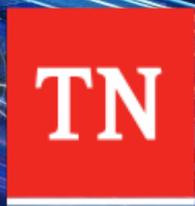

# Connected and Automated Vehicles Investment and Smart Infrastructure in Tennessee

Part 3: Infrastructure and Vehicular communications: From Dedicated Short-Range Communications to Cellular Vehicle-to-Everything

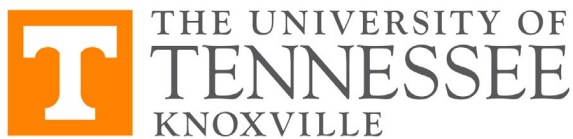

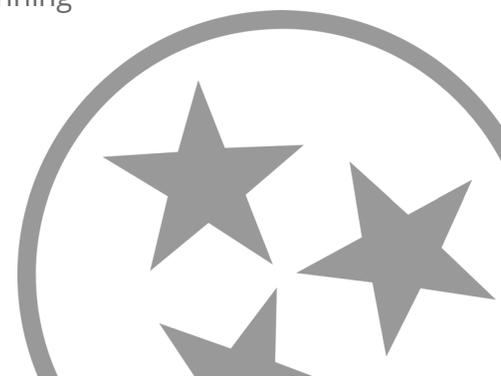

# DISCLAIMER

This research was funded through the State Planning and Research (SPR) Program by the Tennessee Department of Transportation and the Federal Highway Administration under ***RES2019-07: Research on Connected and Automated Vehicles Investment and Smart Infrastructure in Tennessee.***

This document is disseminated under the sponsorship of the Tennessee Department of Transportation and the United States Department of Transportation in the interest of information exchange. The State of Tennessee and the United States Government assume no liability of its contents or use thereof.

The contents of this report reflect the views of the author(s) who are solely responsible for the facts and accuracy of the material presented. The contents do not necessarily reflect the official views of the Tennessee Department of Transportation or the United States Department of Transportation.

# Technical Report Documentation Page

| 1. Report No.<br>RES2019-07 | 2. Government Accession No. | 3. Recipient's Catalog No. | | |
|---|---|---|---|---|
| 4. Title and Subtitle<br><br>*Infrastructure and vehicular communications: From Dedicated Short-Range Communications to Cellular Vehicle-to-Everything* | | 5. Report Date<br>May 2022 | | |
| | | 6. Performing Organization Code<br>R011313595 | | |
| 7. Author(s)<br>Asad Khattak, Austin Harris, Mina Sartipi, Iman Mahdinia, Nastaran Moradloo, & Mohammad SafariTaherkhani | | 8. Performing Organization Report No. | | |
| 9. Performing Organization Name and Address<br>University of Tennessee, Knoxville, Tennessee, 37996<br>University of Tennessee, Chattanooga, Tennessee, 37403 | | 10. Work Unit No. (TRAIS) | | |
| | | 11. Contract or Grant No.<br>RES2019-07 | | |
| 12. Sponsoring Agency Name and Address<br>Tennessee Department of Transportation<br>505 Deaderick Street, Suite 900<br>Nashville, TN 37243 | | 13. Type of Report and Period Covered<br>Final Report.<br>May 2019-May 2022 | | |
| | | 14. Sponsoring Agency Code | | |
| 15. Supplementary Notes<br>The project has developed a series of five (5) reports that support intelligent mobility strategies in Tennessee. This is report 3 of 5. | | | | |
| 16. Abstract<br>This report aims to support the Tennessee Department of Transportation's decisions about vehicle and infrastructure communications technologies. The transition from Dedicated Short-Range Communication (DSRC) V2X to Cellular Vehicle to Everything (C-V2X) is explored using USDOT guidance on relevant issues and presenting the results of experimentation in Tennessee and the potential pros and cons. DSRC V2X technology has been planned at traffic signals in Tennessee, e.g., 152 Roadside Units (RSUs) were planned by TDOT using DSRC V2X and Bluetooth combination units in the I-24 smart corridor. Similarly, many pilot programs and testbeds around the nation have deployed DSRC V2X technology and are now impacted by the Federal Communication Commission's (FCC) ruling on opening safety band. The implication is that existing DSRC V2X deployments (and future deployments) should migrate to C-V2X. Notably, dual-mode RSUs are available along with LTE C-V2X. The transition can be done by working with vendors, but surely this involves more than swapping DSRC V2X devices with LTE C-V2X devices. Complicating the migration to C-V2X is TDOT's role in traffic signal operations and maintenance, which is limited to funding and designing/construction of traffic signals, but local agencies operate and maintain signals. Hence local agencies will work with TDOT to operate and maintain C-V2X technology. Moreover, C-V2X technologies are not widely tested-interference by unlicensed devices and channel congestion can adversely affect safety-critical applications. Given the substantial uncertainties in transitioning to these technologies, TDOT's discussions with IOOs about the operation and maintenance of C-V2X may have to wait for the resolution of issues, while TDOT can invest in experimentation with dual-mode devices. Recommendations are provided about dual-mode devices, CAV data, and needed research and testing. | | | | |
| 17. Key Words<br>**Cellular Vehicle-To-Everything, Dedicated Short-Range Communication, Dual-Mode Devices, Tennessee, Safety** | 18. Distribution Statement<br><br>No restriction. This document is available to the public from the sponsoring agency at the website http://www.tn.gov/. | | | |
| 19. Security Classif. (of this report)<br>Unclassified | 20. Security Classif. (of this page)<br>Unclassified | 21. No. of Pages<br>34 | 22. Price | |

# Acknowledgment

We are very grateful to the Tennessee Department of Transportation staff for their input and particularly to the following individuals for their support throughout the project: Mr. Lee Smith, Mr. Jon Storey, Mr. Brad Freeze, and Ms. Melanie Murphy. Mr. Edward Fok of the Federal Highway Administration, USDOT, provided very valuable information to the research team for which we are very grateful.

Several students have contributed to this report and project. They include Ms. Nastaran Moradloo, Mr. Mohammad SafariTaherkhani, Mr. Latif Patwary, Mr. Steve Lee, Ms. Antora Hoque, and Mr. Zachariah Nelson.

# Executive Summary

## Background

Using the five levels of automation suggested by the Society of Automotive Engineers and the US Department of Transportation (US DOT), the operation of higher-level Connected and Automated Vehicles (CAVs) is enabled by the exchange of data between smart infrastructure and equipped vehicles. A vehicle can also share data with other nearby vehicles. Dedicated short-range communication (DSRC) for vehicle-to-everything (V2X) and cellular vehicle-to-everything (C-V2X) are two technologies used for very high-speed

> *The goals of TDOT's smart infrastructure project are to provide:*
> - *A complete picture of relevant research, development, and deployment (RDD)*
> - *Discuss key research findings and investment opportunities*
> - *Provide recommendations for investments in intelligent mobility*

and high-frequency data transmission (up to 10 times per second with millisecond latency). These technologies can provide travel information, rear-end crash warnings, red-light running notices, slippery roadway information, critical weather warnings, and curve speed alarms. To support the Tennessee Department of Transportation (TDOT) regarding decisions on communications technologies, the research team has compared different standards for vehicle communications (DSRC V2X and C-V2X technologies) and explored the transition from DSRC V2X technology to C-V2X given the recent safety band ruling by the Federal Communications Commission (FCC). Notably, C-V2X and DSRC V2X technologies have similarities, including sharing similar information such as location, acceleration, and speed, utilizing a digital signature, and using the same message sets. However, they have several differences, such as different spatial coverage and ranges, latency and packet drop rates, and communication beds based on their different chipsets. Smart infrastructure deployment in the US will require a successful transition from DSRC V2X to C-V2X communications, especially with roadside units. This transition might be costly and needs substantial time and effort. Despite the advantages and disadvantages of the two technologies (DSRC V2X and C-V2X), comprehensive testing that compares their performance is limited. Moreover, one of the critical aspects of C-V2X technology that encouraged the industry to move forward with the transition to this technology is the compatibility of mobile phone technology and increasingly low latency for safety applications. However, as C-V2X communication technology has not been tested at a large scale, the technology currently has substantial uncertainty, which can be hedged by using dual-mode C-V2X/DSRC V2X devices.

## Key findings

The key findings from a review of recent USDOT guidance on vehicular communication and the literature highlight the following issues.

- **Transition to C-V2X technologies.** Agencies throughout the US that have deployed DSRC V2X are now transitioning to C-V2X technologies, which nevertheless involves uncertainty given the complexities of procuring and installing the software and hardware. Furthermore, technology, equipment, standards, and uses for C-V2X are evolving and agencies have found the transition from DSRC V2X to C-V2X challenging. While the transition undertaken by TDOT can be done by working with vendors, this involves more than swapping DSRC V2X devices with Long-Term Evolution (LTE) C-V2X devices or installing dual-mode devices that can perform both DSRC V2X and C-V2X communications. Notably, these devices and associated software are still in their infancy, have limited availability through vendors, and entail several complicated steps, with associated uncertainties, according to a recent National Cooperative Highway Research Program report [1]. This transition will involve more time and resources.

- ***Issues in C-V2X transition.*** Complicating the migration to C-V2X is TDOT's role in traffic signal operations and maintenance, which is limited to funding and designing/constructing traffic signals. However, local agencies (infrastructure owners and operators-IOOs) operate and maintain signals, and TDOT works and assists local agencies in the smooth operation of signals. Hence local agencies will work with TDOT to deploy, operate, and maintain C-V2X technology. Further complications come from LTE C-V2X interference from unlicensed devices and channel congestion, adversely affecting safety-critical applications. Since there are substantial uncertainties in transitioning to these emerging technologies, discussions with IOOs about the operation and maintenance of C-V2X may have to wait to resolve these issues. At the same time, TDOT can invest in limited experimentation with dual-mode devices. Since there are substantial uncertainties in transitioning to these emerging technologies, IOOs in Tennessee can examine the operation and maintenance of dual-mode C-V2X. Notably, TDOT has recently issued guidance recommending a transition to dual-mode C-V2X devices if IOOs are in planning or design stages. Meanwhile, it will be prudent to develop investment plans to experiment with dual-mode C-V2X devices.

- ***Smart infrastructure deployment.*** In Tennessee, smart infrastructure deployment will require a successful transition from DSRC V2X to C-V2X communications, especially with roadside units, e.g., installed at traffic signals or on freeways. This transition has a high opportunity cost for changing communication devices and delaying the deployment of safety-critical applications through the available DSRC V2X devices. Specifically, researchers at the University of Michigan Transportation Research Institute [2] point out that the cost of delaying the deployment of safety-critical applications through the available DSRC V2X can be measured in terms of tens of thousands of lives lost over five years.

- ***Connected vehicle applications and needed research.*** Connected vehicle applications in Tennessee have relied on the DSRC V2X communication platform for Transit Signal Priority and Emergency Vehicle Preemption applications. CAV data collection, use, security, and storage are critical aspects going forward. Notably, 4G LTE-based C-V2X communication technology is being tested by the US DOT. Early results of data analysis indicate that there can be congestion issues during operation compared with DSRC V2X. However, presently there is not enough evidence to confirm or contradict whether LTE C-V2X will scale up to the safety issues. More research is needed on several aspects of communication technology. While the FCC has provided a transition plan for moving from DSRC V2X into the new C-V2X spectrum, the shift will nevertheless be costly.

## Key Recommendations

The recommendations regarding V2X communications are structured around three main themes, the transition, data, and research/testing:

- ***Managing the transition to dual-mode C-V2X and DSRC V2X devices.*** Given that C-V2X technology is still being examined for large-scale deployment, the transition will include testing and using dual-mode DSRC V2X and C-V2X radio devices. Furthermore, strategies can include investing in the reliable installation of software updates for C-V2X (e.g., updates that address security vulnerabilities within signal controllers, so they can communicate securely with C-V2X devices in the future); recruiting volunteers from the general public and fleet vehicles (owned by TDOT, or other agencies) for installing onboard units; and exploring automotive/Original Equipment Manufacturer (OEM) industry's acceptance of C-V2X technologies. Importantly, experimentation with dual-mode devices by the University of Tennessee, Chattanooga, has shown promising results for dual-mode devices. Specifically, experimentation in the Chattanooga MLK smart corridor shows that LTE 4G provides stable performance (than DSRC V2X) when considering packet loss and range in an urban environment.

- ***Managing and harnessing connected and automated vehicle data.*** It is vital to identify CAV data needs and types (e.g., basic safety messages, SPaT messages, and vehicle trajectories). Equally important are data analytics and cybersecurity investments for streaming data generated by CAVs. Data management plans are also needed. Harnessing the data is an opportunity to improve traffic operations and performance evaluation. TDOT, with private sector partners, can collect high-resolution data from traffic signals and vehicle probe data, allowing engineers to track traffic trends at an intersection, corridor, and programmatic level as well as visualize maintenance problems. In this regard, TDOT can use Automated Traffic Signal Performance Measures (ATSPM) software to oversee real-time and historic functionality at individual intersections. All intersections communicating with TDOT can be configured in ATSPM. ATSPM can find faults and errors, thereby saving staff field inspection time. This software has several use cases, including Utah DOT, Virginia DOT, and Georgia DOT. In addition, TDOT can consider using software named TEAMS (Traffic Engineering and Asset Management Software) to track traffic signal equipment, including signs, Americans with Disabilities (ADA) compliance, and overall rating of intersections. For equipment maintenance updates, if the equipment is not performing correctly, it is noted as a task in TEAMS and updated as needed.

- ***Future research on communication technologies.*** Future research can entail testing the performance in terms of latency and packet drop of C-V2X technologies and the impacts of (user) congestion and interference in communication networks. Also, it is important to assess TDOT's tolerance for latency and packet loss in safety-critical applications. Furthermore, investments in addressing the uncertainty of C-V2X in terms of the scale (number of vehicles) will be valuable, along with testing the longevity and compatibility of C-V2X technology performance in different environmental conditions within Tennessee. Noting that the current 30 MHz for communication may not be sufficient, there is a need to investigate how the 5.9 GHz communication band can support C-V2X. More generally, for TDOT to deploy C-V2X technologies, transportation system readiness should be assessed. Overall, substantial and comprehensive testing and research are needed before fully transitioning to C-V2X.

# Table of Contents

# List of Tables

# List of Figures

# Glossary of Key Terms and Acronyms

**ADAS** - Advanced Driver Assistance Systems
**BSW**-Blind Spot Warning
**CAV** – Connected and Automated Vehicle
**CDOT** - Chattanooga Department of Transportation
**CMAQ** - Congestion Management and Air Quality
**CV** - Connected Vehicle
**C-V2X** - Cellular Vehicle-to-Everything
**DMP** -data management plan
**DSRC** - Dedicated Short-Range Communication
**EEBL**-Electronic Emergency Brake Light
**EV** – Electric Vehicle
**FCC** – Federal Communication Commission
**FCW** - Forward Collision Warning
**FHWA** - Federal Highway Administration
**GNSS** - Global Navigation Satellite System
**HIP** - Highway Infrastructure Program
**HSIP** - Highway Safety Improvement Program
**IMA** - Intersection Movement Assist
**IOO** - Infrastructure Owner-Operators
**ITS** - Intelligent Transportation System
**LED** - Light-Emitting Diode
**LTE** – Long-Term Evolution
**NHPP** - National Highway Performance Program
**NHTSA** - National Highway Traffic Safety Administration
**OBU** – Onboard Unit
**OEM** - Original Equipment Manufacturer
**OTA** - Over-The-Air
**RDS** - Radar Detection System
**REL** - Reversible Express Lane
**RFP** - Request for Proposals
**RR** - Refuge Roads
**RSU** – Roadside Unit
**SAE** - Society of Automotive Engineers
**SCMS** - Security Credential Management System
**SPaT** - Signal Phasing and Timing
**STBG** - Surface Transportation Block Grant
**TDOT** – Tennessee Department of Transportation
**TEAMS** - Traffic Engineering and Asset Management Software
**THEA**-Tampa Hillsborough Expressway Authority
**TMC** - Traffic Management Center
**UM** - University of Michigan
**UTC** - University of Tennessee at Chattanooga
**UTK** – University of Tennessee at Knoxville

**US DOE** - US Department of Energy
**US DOT** - US Department of Transportation
**V2I** - Vehicle-to-Infrastructure
**V2V** – Vehicle-to-Vehicle
**V2X** – Vehicle-to-Everything
**WAS** - Worker Alert System
**WAVE** - Wireless Access in Vehicular Environments

# Chapter 1  Introduction

Connected and Automated Vehicles (CAVs) are becoming more widely tested and deployed, e.g., pilot tests in the US include the Tampa Connected Vehicle (CV) Pilot, New York CV pilot, and Wyoming CV pilot. Several states are creating standards and guidelines for CAVs that span planning, infrastructure upgrades (upgraded

*This report aims to synthesize guidance for the deployment of connected vehicle technology, which is moving from DSRC V2X to C-V2X.*

traffic signal controllers and communication technologies), and incorporating communication and connectivity technologies into the highway network. To support the Tennessee Department of Transportation's (TDOT) infrastructure investment decisions, this report compares different standards for vehicle communications and the transition from Dedicated Short-Range Communication (DSRC) Vehicle-to-Everything (V2X) technology to Cellular Vehicle-to-Everything (C-V2X), motivated by the Federal Communications Commission (FCC) ruling on the safety band. US Department of Transportation (US DOT) guidance on relevant issues is explored along with these technologies' potential pros and cons. Furthermore, the report provides information about CAV data as well as needed research and testing of communications technology

## 1.1 Research Objectives

The objective of this report is to synthesize guidance for the deployment of connected vehicle technology, which is (inevitably) moving from DSRC V2X to C-V2X, typically installed at traffic signals within the state of Tennessee and in some cases along freeways or expressways. For example, of the 152 roadside units (RSUs), 30 will be installed by TDOT along the I-24 Smart Corridor using DSRC V2X and Bluetooth combination units, and 122 RSUs are planned along State Route 1/surface streets by local agencies, who will also have responsibility after the units are operational. Similarly, many pilot programs and testbeds around the nation have deployed DSRC V2X technology and are now impacted by the FCC's ruling on opening safety band allocation. Specifically, the FCC repurposed 45 megahertz of the 5.850-5.925 GHz band (the 5.9 GHz band) spectrum to enable expansion of unlicensed mid-band spectrum operations [3], while continuing to dedicate only 30-megahertz spectrum for vital intelligent transportation system (ITS) operations. Further, the FCC claimed that to promote the most efficient and effective use of the ITS spectrum, they require the ITS service to use C-V2X-based technology at the end of a transition period. The FCC further claimed that "by splitting the 5.9 GHz band between unlicensed and ITS uses, [FCC's] decision puts the 5.9 GHz band in the best position to serve the needs of the American public." The FCC also mentions that "DSRC has barely been deployed, in the more than 20 years since the adoption of DSRC, meaning this spectrum has been largely unused [3]." The implication is that existing DSRC deployments (and future deployments) should plan to migrate to C-V2X.

The safety band is the wireless spectrum reserved for transportation-related communications among the devices that support connected and automated vehicles. Until recently, an interference-free safety band was available, enabling communications between vehicles and traffic signals, helping generate real-time alerts or warnings, and adjusting signals to give emergency vehicles priority in heavy traffic. However, congestion and interference are a concern in the future, given the FCC spectrum reallocation decision. This report provides information and experimentation to support TDOT decisions about CAV technology deployment and the developments of the safety band. To this end,

- The University of Tennessee at Chattanooga (UTC) has performed preliminary testing of emerging dual-mode devices (that offer both DSRC V2X and C-V2X) devices, with the results provided in this report.
- The University of Tennessee, Knoxville (UTK) team has explored federal guidance on the safety band issue and synthesized projects in the US in which DSRC V2X or C-V2X communication technologies are being deployed and tested. Helpful information is extracted and organized in lessons learned from these projects and recommendations for TDOT about transitioning from DSRC V2X to C-V2X.

## *1.2 Organization of the Report*

The report is organized into the following sections:

**Chapter 2 – Methodology and Strategy for Transitioning from DSRC V2X to C-V2X.** This chapter discusses the methodology used in this study based on a review of relevant literature, reports, and websites, experimentation with dual-mode C-V2X and DSRC V2X devices, and interviews.

**Chapter 3 – Guidance on the Installation of OBUs and RSUs.** This chapter summarizes the guidance on installing onboard units (OBUs) and RSUs in Tennessee.

**Chapter 4 – Experimentation: MLK Smart Corridor Chattanooga.** This chapter presents the guidelines for preparing the CAV strategic plan in Tennessee.

**Chapter 5 – CV Pilot Successes and Lessons Learned.** This chapter provides details about the successes and lessons learned about communication technologies from CV pilot programs.

**Chapter 6 – Illustrative Practices in Other States.** This chapter discusses case studies that illustrate how states are procuring RSUs and OBUs from vendors.

**Chapter 7 – Conclusions and recommendations.** The findings are summarized along with the contributions of the reported work. A discussion of recommendations is provided.

# Chapter 2 Methodology and Strategy for Transitioning from DSRC V2X to C-V2X

The methodology used in this study is based on a review of relevant literature, reports, and websites; experimentation with dual-mode C-V2X and DSRC V2X devices in the MLK smart corridor testbed; and an interview and input received from Mr. Edward Fok of the Federal Highway Administration (FHWA), US DOT.

*For TDOT, transition to C-V2X can entail:*
- *Deploying dual-mode DSRC V2X and LTE- or 5G-V2X devices*
- *Waiting for the results and findings of the national, state, and private-sector studies*
- *Testing DSRC V2X, LTE-V2X, and 5G-V2X technology on an experimental basis*

While the transition from DSRC V2X to C-V2X seems imminent in the US, it is worthwhile to consider their performance and status. The current testing of C-V2X systems is based on 4G Long-Term Evolution (LTE)-based cellular V2X. A summary of the US DOT's ongoing study based on scenarios to examine the LTE-V2X radio performance in safety-critical transportation conditions is presented in Appendix A. The LTE-V2X tests generated large amounts (terabytes) of data that are being analyzed to provide insights into the ambiguities associated with C-V2X technology, e.g., whether it will work effectively when the scale (the number of vehicles) increases and whether it will improve vehicle safety in complex situations. The DSRC V2X scale is better known and tested. However, the scale of the LTE V2X technology is one area that requires further research and investigation. Early results show that there can be congestion issues compared with DSRC V2X. Many questions need to be answered about transitioning to C-V2X, which will be discussed in this section.

Notably, C-V2X is a new technology that requires more time, research, and investigation to be ready for large-scale deployment. Therefore, at this time, strategies that TDOT can consider for transitioning to C-V2X, given the uncertainty associated with C-V2X technologies, include:

1. Deploy **dual-mode** DSRC V2X and LTE- or 5G-V2X devices that can be upgraded based on future advancement.
2. **Wait** for the results and findings of the national, state, and private-sector studies to pave the road for C-V2X to be ready for large-scale deployment.
3. Deploy DSRC V2X, LTE-V2X, and 5G-V2X technology on an experimental basis and **collaborate** with the US DOT, state Departments of Transportation (DOTs), and private sector vendors and test the technologies. This will entail choosing the best option for TDOT's transportation needs, e.g., required latency and packet drop thresholds.

Some of these options are not necessarily mutually exclusive. Relevant and insightful information about each of these technologies, including their strengths and weaknesses, is provided in this chapter to help TDOT select the best strategy. More relevant information can be found at the following website:

https://www.ite.org/technical-resources/standards/connected-intersections/

## 2.1 DSRC V2X versus C-V2X: Comparing the Connected Vehicles Technologies

The development and deployment of DSRC V2X technologies are considered a substantial improvement by agencies and the automotive industry as it enables data to be transmitted without going through any intermediaries. As a result, DSRC V2X is very useful for areas without any telecommunication infrastructure, such as rural areas. Additionally, DSRC V2X has very low latency based on eliminating intermediaries. C-V2X uses cellular radio (similar to cellphones) rather than WLAN utilized in DSRC V2X. An important difference between DSRC V2X and C-V2X is that C-V2X enables direct and indirect messages, while DSRC V2X only provides direct messages. In fact, in the direct C-V2X, vehicles communicate directly with other vehicles (V2V) and roadside units (V2I) the same way DSRC V2X can communicate. On the other hand, in the indirect C-V2X, vehicles indirectly communicate with other entities via cellular networks (V2N), which DSRC V2X cannot do. This enables real-time vehicle information to be provided to a traffic control center, which can be harnessed for operational improvements [4] [5].

As a result, both technologies allow vehicles to communicate with each other (V2V) and with surrounding infrastructure (V2I). However, a DSRC V2X device using a WAVE wireless standard cannot communicate with a C-V2X device using LTE. This lack of interoperability has resulted in a fragmented market, and different CV deployments have selected one of these technologies for implementation. Only recently dual-mode devices have become available.

Internationally, countries and regions hold different views about transitioning to C-V2X from DSRC V2X, which have evolved. It is notable that, similar to the US, DSRC V2X-based V2X communication was deployed in Europe and Japan until recently. However, most counties and regions are transitioning to C-V2X, some slower than others. Notably, China has moved most aggressively toward the deployment of C-V2X technologies, as this new technology is viewed as effectively addressing China's safety and congestion issues. Additionally, automobile manufacturers and their suppliers seem fragmented about which technology to use for V2X connectivity. Specifically, Toyota, General Motors, and Volkswagen have indicated a preference for DSRC V2X-equipped automobiles, while Daimler, Ford, and Stellantis (14 brands including Chrysler, Fiat, and Jeep) have endorsed the cellular-based V2X [6].

Based on a connected and automated vehicles survey conducted in 2019, information from 258 agencies located throughout the United States that were deploying or planning to deploy CVs [7], a majority were using DSRC V2X compared with CV2X technology (Figure ). However, it is notable that quite a few agencies were using C-V2X technology, especially public transit agencies.

The US DOT-sponsored CV pilots also used DSRC V2X technology. Deployment of DSRC V2X is reflected in:

- TDOT's signature project-the I-24 Smart Corridor and the MLK Smart Corridor in Chattanooga, use DSRC V2X technology for various applications.
- The Tampa, FL CV pilot uses the Pedestrian Collision Warning Application technology.
- The Wyoming CV pilot uses the technology on Interstate-80 for communication with commercial vehicles.
- The NYC CV pilot uses the technology at signalized intersections to improve performance.

- The Georgia DOT uses DSRC V2X for signalized intersections, according to the statewide CV Deployment Plan.

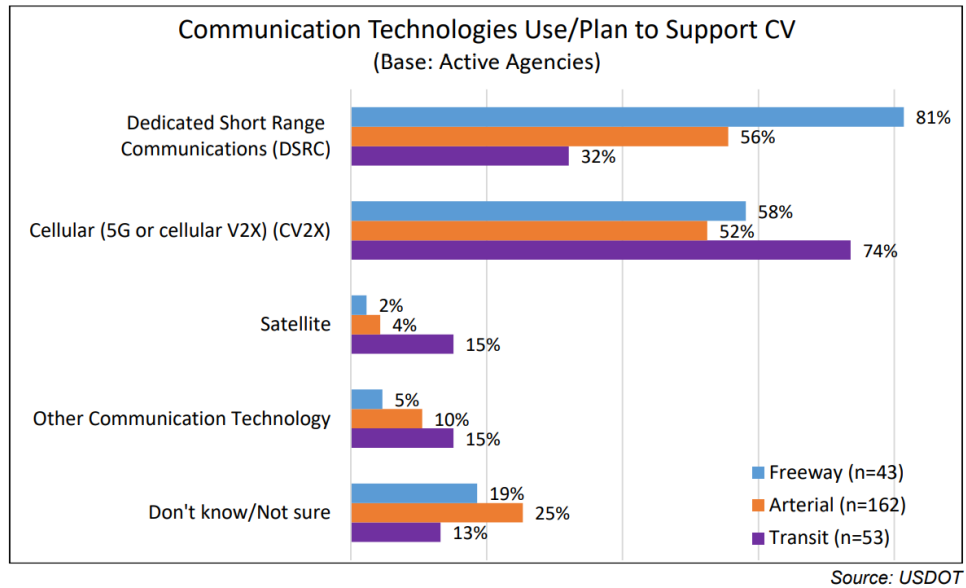

**Figure 2-1 Percentage of different technologies that active agencies were using in 2019**

**Source:** https://www.itskrs.its.dot.gov/sites/default/files/doc/2019_connected_vehicle_automated_vehicle_survey.pdf

Notably, DSRC V2X uses high-speed communication even when there are physical obstacles or extreme weather conditions. The technology can operate at very high driving speeds, up to 500 km/h (310 mph). The range of DSRC V2X is usually 300 meters (about 1,000 ft), but studies suggest the range can go wider if needed. C-V2X has a similar range of around three hundred meters, perhaps 20% to 30% wider. Similar results were observed in the preliminary work conducted by UTC. Overall, it is possible that C-V2X capabilities could handle 1 million connected devices per square kilometer.

One of the major benefits of DSRC V2X is that it has already been designed, verified, and deployed in regional testing pilots in the United States. C-V2X, while not yet being used at the same scale, offers advantages to various smart city and Internet of Things goals that may result in cities making investments in this technology for multiple uses. This transition (partly motivated by the FCC rulings) to C-V2X in connected vehicles can be more manageable.

## *2.2 Similarities and Differences between DSRC V2X and C-V2X*

Both DSRC V2X and C-V2X have low latency, with each technology meeting the safety requirements for V2X communication. The technologies feature latency within milliseconds (Figure 2-2). The mean latency of DSRC V2X is low in terms of milliseconds (ms), compared with the latency range of LTE 4, with DSRC V2X representing a feasible and tested technology for V2X communication in safety-critical applications. However, 4G operate in a "yellow" (or gray) latency range, creating uncertainty about their handling of safety-critical applications.

No connection is maintained between devices with either DSRC V2X or C-V2X. Both share the same information; each device broadcasts status elements such as location, acceleration, and speed while listening to other nearby devices simultaneously. Digital signatures are utilized in both technologies to guarantee security and ensure the validity of transmissions. Both C-V2X (4G LTE) and DSRC V2X communicate directly using the 5.9 GHz band and even use the same message sets, Society of Automotive Engineers (SAE) J2945 and J2735 [4]. Similar to 4G, 5G is even lower latency and as 5G cellular services are deployed, most vehicles will utilize these higher speed and bandwidth links as well. However, 5G coverage will diffuse slowly to rural areas compared with urban areas [8], [4].

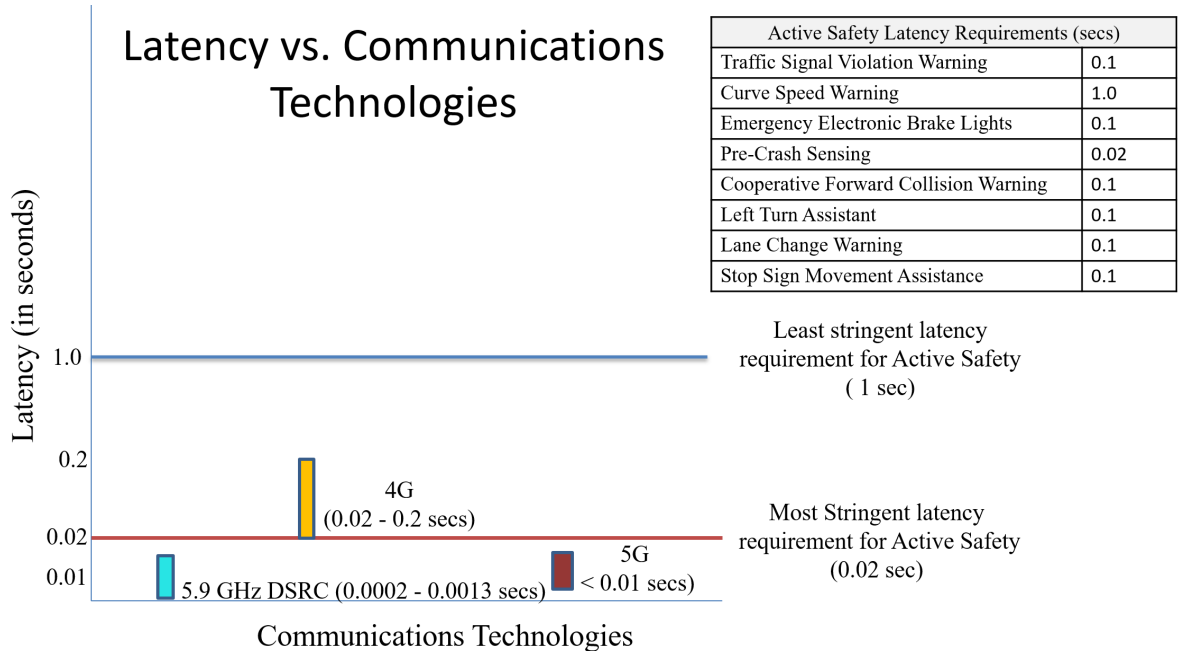

| Active Safety Latency Requirements (secs) | |
|---|---|
| Traffic Signal Violation Warning | 0.1 |
| Curve Speed Warning | 1.0 |
| Emergency Electronic Brake Lights | 0.1 |
| Pre-Crash Sensing | 0.02 |
| Cooperative Forward Collision Warning | 0.1 |
| Left Turn Assistant | 0.1 |
| Lane Change Warning | 0.1 |
| Stop Sign Movement Assistance | 0.1 |

Note: Y-axis not to scale for illustration purposes

**Figure 2-2 Latency vs. communications technologies**

## 2.2.1 Strengths of DSRC V2X Over C-V2X

A crucial advantage of DSRC V2X over C-V2X is time and readiness. Notably, DSRC V2X technology entered production in some Toyota car models, specifically in Japan in 2015, before Cadillac followed suit in 2017. Europe's most popular automobile Volkswagen Golf began utilizing the technology in its eighth-generation model back in 2019. Another key advantage of DSRC V2X is that it has been designed and tested in major CV pilots, including New York (dense urban), Florida (suburban-urban), and Wyoming (rural) environments. This mass-market application is a sign of confidence in DSRC V2X's application. By comparison, C-V2X has only had small initial deployments starting in 2021. DSRC V2X is the more proven technology, already fully designed for the automotive industry and ready for application without relying on existing infrastructure. The Wi-Fi-based wireless DSRC V2X is low latency and optimized for high mobility applications even in the presence of obstructions. DSRC V2X can handle a fast-changing environment at speeds as high as 500 km/h with a communication range of up to 1 km. These features make DSRC V2X feasible for fast exchange of data in passenger vehicles, as well as other modes, such as trains.

Of course, there are drawbacks inherent to wireless communication, such as limited channels, particularly in metropolitan areas, and limitations on the volume of data that can be transmitted and drops of packets. Notably, DSRC V2X messages typically require low data volumes, whereas an automated vehicle using C-V2X technology can use a large amount of data per day. Additionally, DSRC V2X is loose asynchronous while C-V2X has very tight synchronous requirements. DSRC V2X's symbol duration is eight uS fast channel tracking, while C-V2X tracks at around 70 uS. Coding gain for C-V2X uses turbo code and HARQ compared to DSRC's convolution code.

## 2.2.2 C-V2X Strengths Over DSRC V2X

Remarkably, C-V2X (LTE 4G or later 5G) is a potential game-changer for the connected vehicles industry. Because LTE chip technology is embedded in almost all cellular devices, C-V2X can use existing cellular infrastructure, both on the roadside and increasingly found in vehicles, reducing the amount of equipment needed to be installed and maintained. Utilizing already well-established LTE technology can create a comprehensive network, enabling high-speed communication in high-density locations, using technology people are already familiar with.

LTE has a clear upgrade path with 5G networks increasingly becoming available for use, making C-V2X future-facing and taking advantage of new high speeds. In 2021, C-V2X was named the new standard for connected vehicle technology over DSRC V2X in the United States by the FCC in a controversial ruling. The Intelligent Transport System, of which V2X is a part, was limited to using only the upper 30 Megahertz of the 5.9 GHz communication band, making C-V2X the go-to technology for a large market (Figure ) [3].

C-V2X is still in the early stages of application, but tests indicate that it could achieve a 20-30% greater range than its DSRC V2X and even perform better when faced with obstructions. C-V2X, with its SC-FDM modulation, can also achieve concurrent transmissions, while DSRC V2X is not able to do so. The cellular technology overcomes network collisions using retransmission, usually activated at high speed, increasing its communication range [9]. Furthermore, C-V2X increases the reaction time compared to DSRC V2X (Figure ).

Proposed Band plan

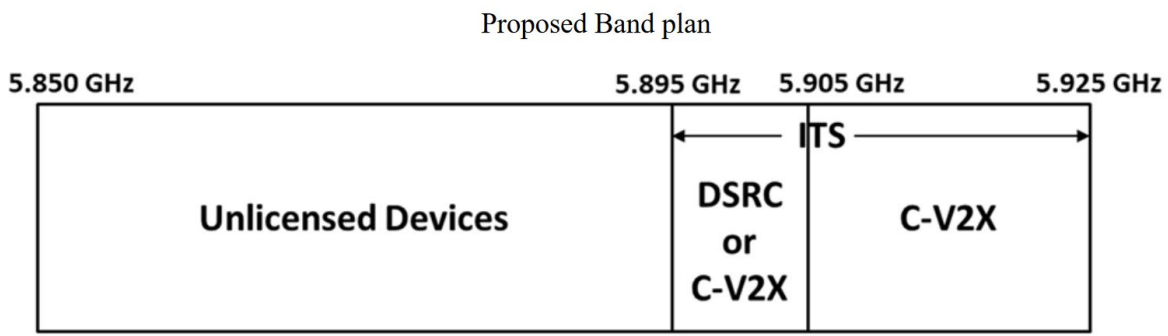

**Figure 2-3 FCC's proposed band plan [3]**

**Source:** https://docs.fcc.gov/public/attachments/FCC-20-164A1.pdf

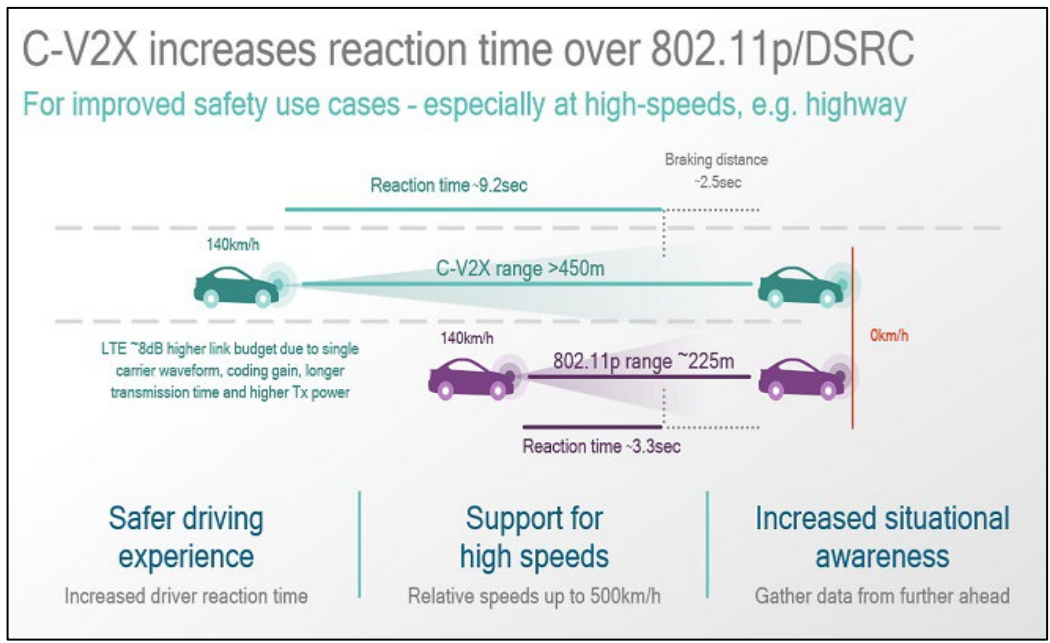

**Figure 2-4 Strength of C-V2X over DSRC V2X for Safety improvement**

**Source:** https://www.ednasia.com/using-dsrc-as-a-v2x-technology/

The Safety Band is a wireless spectrum at 5.9 GHz, and it enables communications among devices that support CAVs. The recent FCC decisions had influenced the safety band when the FCC reallocated 45 MHz of the 5.9 GHz safety band to expand unlicensed mid-band spectrum operations (unlicensed Wi-Fi uses). The FCC has continued to allocate 30 MHz of spectrum for vital ITS operations (i.e., transportation purposes). The new FCC rules also include the following changes:

1. To utilize the spectrum allocated to ITS, the services have to migrate cellular vehicle-to-everything based technology.
2. The "guard band" between the lower end of the original Safety Band and the unlicensed Wi-Fi band was removed, which resulted in public concerns about potential out-of-band emissions (OOBE) from unlicensed Wi-Fi operating just below the new V2X band.
3. The new rules removed the priority for the public safety message, raising concerns about how unlicensed Wi-Fi transmissions may influence such messages.
4. V2X communications must operate at the upper end of the original Safety Band based on changes. Therefore, adjacent band interference from unlicensed devices running above the V2X band can interfere with V2X communications.
5. DSRC V2X users have one year from the date of the first FCC report to continue operating in the entire 75 MHz DSRC V2X users must move to the upper 30 MHz of the band within 12 months from the date of the second Report and Order. Then DSRC V2X must switch to LTE-V2X.

Notably, safety is the main goal of the US DOT. As a result, the FCC's new rules should be evaluated along with safety considerations. Therefore, the US DOT's 5.9 GHz Spectrum Team is examining the radio performance and communication quality of the LTE- C-V2X technology under the new FCC rules in a real-world environment that provides both typical and challenging crash avoidance situations [10]. In addition, the US DOT is assessing the performance of LTE-V2X technology in a dense traffic environment with over 1000 vehicles. Figure shows the required tests to evaluate LTE-V2X communications under varying conditions [9].

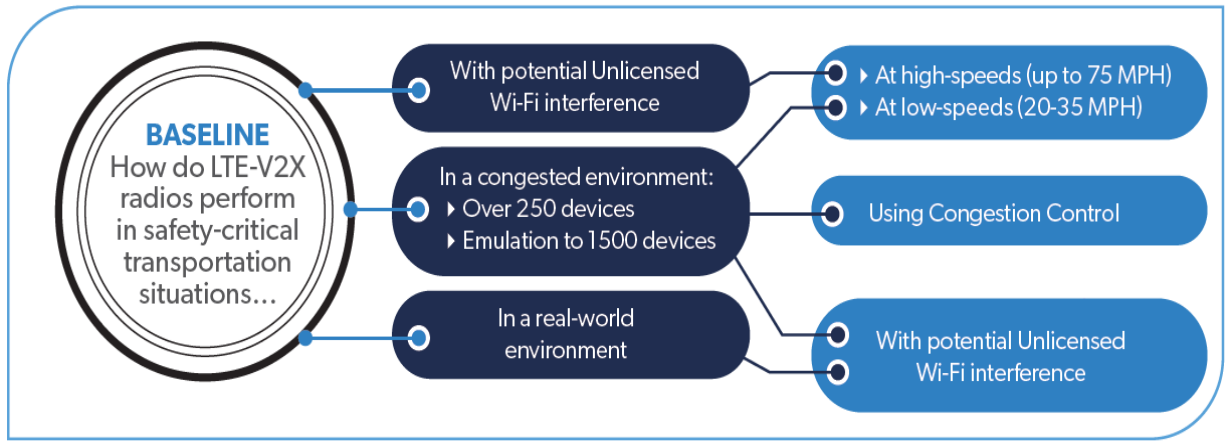

**Figure 2-5 Testing LTE-V2X communications under varying conditions [9]**

## 2.3 The spectrum issue: Questions on Transitioning to C-V2X

To deal with the spectrum issue, i.e., the decision by the FCC on the Safety Band, the US DOT has considered transitioning from the default DSRC V2X to C-V2X. Some questions still need attention for the transition to C-V2X [11]:

***What is the direction of the current, ongoing deployments designed with DSRC V2X?***

DSRC V2X deployments are currently on hold, while some have started transitioning to dual-mode communications. TDOT can wait for the results of ongoing C-V2X research and testing projects before embracing this technology. The federally sponsored research should provide guidance and insight for decisions on the transition to large-scale deployment of C-V2X. Currently installed DSRC V2X-based operational equipment, if present, can be retained, given their value in terms of operational performance and substantial uncertainties associated with the transition to C-V2X.

***How should transition to C-V2X be navigated to implement large-scale, city-wide V2X deployment infrastructure?***

Switching to C-V2X requires a significant amount of effort. The changes can be navigated by considering the requirements for the performance needed (e.g., in terms of latency and packet drop) for safety-critical applications and hedging by experimenting with dual-mode devices. The costs of technology devices themselves may be relatively low.

***Do driverless vehicles need DSRC V2X or C-V2X technology?***

Driverless vehicles refer to higher levels of automation, i.e., SAE Level 3 and higher. Highly automated vehicles connected with the infrastructure will require a substantial amount of data transfer with the infrastructure, e.g., traffic signals. This raises an important issue: TDOT does not own or operate traffic signals or currently has access to data lakes or servers to collect such data. While TDOT can implement higher-level automation on freeways and expressways, and plan for what the Traffic Management Centers (TMCs) would need to deploy appropriate technology to support higher levels of automation, traffic operations on city streets and especially at traffic signals will remain a complex issue. Wide-scale adoption of higher automation levels must be addressed through partnerships and in the context of increasing CAV readiness. Specific questions to be answered include responsibility for uploading and maintaining the MAP message set (developed under the Society of Automotive Engineers, SAE J2735 standard) guidance, ensuring the Signal Phasing and Timing (SPaT) messages are working accurately, and collecting and processing the relevant data streams. Given that TDOT does not own or operate traffic signals and the backhaul communications or store CAV data, the best course of action is to partner with local agencies that own and operate traffic signals and roadways in urban areas. Also, where appropriate, TDOT can use regional TMCs to coordinate with city TMCs in terms of their signal operations. Furthermore, academic institutions or other entities can also assist with developing and deploying CAVs, e.g., serve as repositories of data. Other options are discussed elsewhere in the project reports.

To ease the in-vehicle computational burden, highly automated vehicles will likely provide additional information to infrastructure owner-operators and TDOT, such as 3D LiDAR or 3D video snapshots. While standards are available for 3D imagery and LiDAR point clouds, such data is large and cannot be shared in real-time based on existing communication media. Notably, 5G cellular communication may provide such data exchange from CAVs to IOOs and TDOT, which C-V2X can support in the future. Therefore, in the future, when fully connected vehicles diffuse through the transportation system, DSRC V2X communication will not be able to support such vehicles, and there will be a need for 5G C-V2X technology. Driverless vehicles will most likely need 5G C-V2X technology.

***Will funding be available to transition to LTE C-V2X?***

Typically, allocation of funding will be an internal decision for TDOT. Congestion Management and Air Quality (CMAQ) funding can be potentially used to transition to C-V2X and their wider deployment. The Infrastructure Investment and Jobs Act (IIJA) includes $274 billion in new federal spending on transportation. In the coming years, Tennessee will be receiving $6.2 billion (2021 to 2026) through IIJA, which can be a critical source of funds for the transition and broader deployment of C-V2X technology. The act encourages investments in new and emerging technologies and smart infrastructure and has funding for broadband deployment.

One aspect that needs to be considered is retrofitting stranded DSRC V2X assets eligible for transition to C-V2X technology. The IIJA allocates a portion of the funds per year to this part [12]. However, the funding allocation process generally does not readily permit funding new items until the new product outperforms the existing (default) product in terms of performance criteria. Notably, there is still uncertainty about C-V2X performance for safety-critical applications, and congestion and interference issues, highlighting the need to proceed carefully when funding the transition to C-V2X. Having said this, niche areas to explore include deployment in areas with a high proportion of disadvantaged and under-represented groups and would greatly benefit from CAV accessibility. The sources of funds for intelligent mobility can include core programs (NHPP-/National Highway Performance Program, STBG-Surface Transportation Block Grant, HSIP-Highway Safety Improvement Program, RR-Refuge Roads, CMAQ, and Freight), as well as new programs related to the resiliency of infrastructure, i.e., adapting the transportation system to achieve greater infrastructure resilience with smart technologies. For Tennessee, there is new funding of $158 million for Promoting Resilient Operations for Transformative, Efficient, and Cost-saving Transportation (PROTECT), which can be a new source of funds. Notably, some of the IIJA funds are grants that will need resources to prepare proposals and plans.

**Will the US DOT/NHTSA reconsider the V2V regulations? (There are concerns about the next steps for equipping fleet vehicles with V2X communication.)**

The National Highway Traffic Safety Administration (NHTSA) evaluates advanced driver assistance systems (ADAS) in vehicles, e.g., lane departure warning, forward collision warning, and crash imminent braking. NHTSA's New Car Assessment Program tests new vehicles and rates them on occupant protection in crashes, and vehicles are assigned star ratings. Given the rapid changes in communication technologies, updates to NHTSA's new car assessment program can be expected. These updates can encourage the adoption of V2X technologies that are most beneficial. Therefore, state DOTs observing NHTSA updates regarding communication technologies (in addition to communication with local automobile manufacturers and Original Equipment Manufacturers (OEMs)) can help know the direction of the adoption of communication technologies by the industry.

**Will 30 MHz be enough to support C-V2X in the future?**

The current 30 MHz may not be enough to support the complete need for C-V2X in the future for a cooperative and automated transportation environment, truck platooning, or cooperative maneuvering of automated vehicles. Additional research is needed to quantify the actual needs of the 5.9 GHz communication band for supporting C-V2X. State transportation agencies await FCC rules for the safety band in the future (Second Report and Order). This can further clarify the decision about issues in transitioning to C-V2X.

**Is there any solution for preventing the interference issues between C-V2X and unlicensed use in other areas of the spectrum (Figure )?**

Analysis shows that transportation safety can be affected due to significant adjacent channel interference between the different radio services, including DSRC V2X, LTE C-V2X, and unlicensed Wi-Fi (UNII) users [13]. In this situation, FCC's guidance in the future can help with the final Wi-Fi parameters.

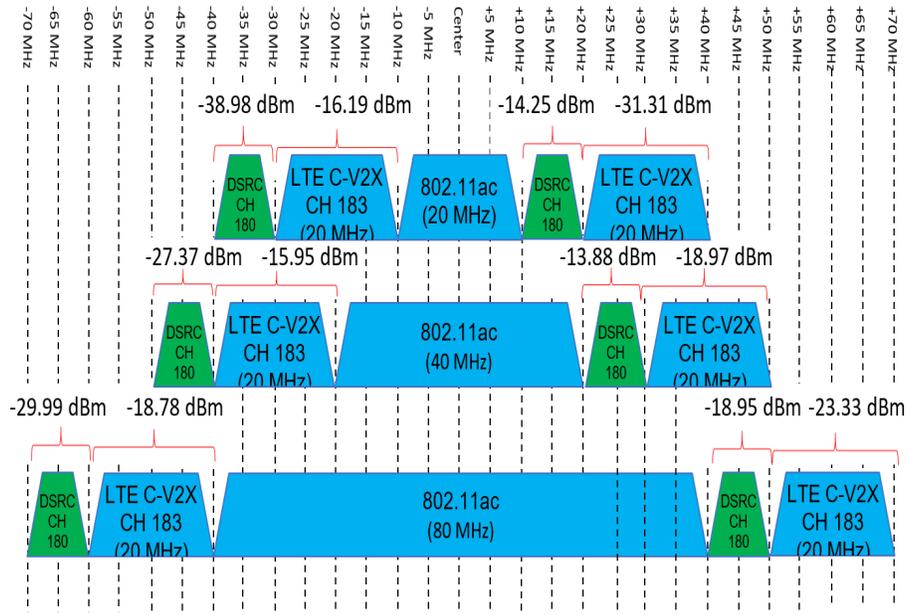

**Figure 2-6 Potential interference between C-V2X and DSRC V2X [11]**

# Chapter 3  Guidance on the Installation of OBUs and RSUs

In September 2016, FHWA completed a project that included the process of deploying DSRC V2X systems. As part of this project, they provide the V2I Hub deployment checklist and guidance documents, including design, approval, and installation activities required to implement the deployment of DSRC V2X systems successfully. The DSRC V2X RSU specification document provides details on hardware/software requirements that would be utilized to support SPaT, including power, environmental condition,

*C-V2X is still an experimental technology and substantial support from vendors is needed. Nevertheless, many state DOTs, universities and research labs, private sector industry, and CV pilots utilize dual-mode DSRC V2X coupled with C-V2X on the roadside and onboard units.*

physical condition, performance, behavioral circumstance, and interface. The previous specification for DSRC V2X-based RSU was based on RSU spec 4.1, which the connected vehicle pilot used. Now it is ITE RSU version 1. This includes a companion portion that connects to the intersection standards. It is a guideline on consistently generating the intersection map and converting the signal timing to signal timing messages. DSRC V2X RSU should work if guidelines are followed. However, actual installation in the New York City CV pilot encountered difficulties installing aftermarket OBUs. Wyoming CV pilot has deployed about 400 OBUs which use DSRC V2X. They have the functionality to broadcast Basic Safety Messages Part I and can include a human-machine interface to share alerts and advisories to drivers of these vehicles [8].

As noted earlier, C-V2X is an experimental technology and there is no concrete recommendation currently about widespread deployment. Substantial support from vendors will be needed. Nevertheless, many state DOTs, universities and research labs, private sector industry, and CV pilots utilize dual-mode (DSRC V2X and C-V2X) RSUs and OBUs. The THEA CV pilot in Tampa, FL, builds on the project's existing DSRC V2X technology by deploying RSUs with dual-mode capabilities. The University of Hawaii and the Hawaii DOT deployed a 5-mile long arterial with dual-radio DSRC V2X/C-V2X connected vehicle technology at 36 traffic signals in Honolulu. The City of Alpharetta, GA, utilized dual-mode DSRC V2X/C-V2X radio technology to alert drivers when entering school zones or near school buses.

# Chapter 4  Experimentation:  MLK  Smart Corridor Chattanooga

The MLK Corridor, an urban testbed in Chattanooga (Figure 4-1), will provide experimental results and insights about the deployment of C-V2X technologies and test their efficacy in real-time under real-world conditions. The specific tested parameters include delay, percentage of packets received, throughput, and coverage. UTC has recently collected preliminary data used to evaluate the performance of both C-V2X and DSRC V2X for these parameters. The UTC team has outfitted a vehicle with an OBU to collect data broadcast via DSRC V2X and C-V2X simultaneously.

> *Based on the findings, DSRC V2X shows a higher packet-loss percentage with a lower transmission range than C-V2X. Moreover, the results show that when the line-of-sight reduces in the test area, it significantly reduces the range for DSRC V2X while having a less significant effect on C-V2X. The maximum range recorded based on findings is 427 meters and 514 for DSRC V2X and C-V2X, respectively.*

Ideally, the mmWave technology can be used with next-generation wireless communications, i.e., 5G, in safety-critical CAV applications. Vendors are increasingly providing new dual-mode V2X systems featuring software and hardware that enables DSRC V2X and cellular (LTE or preferably 5G) C-V2X communication. Accordingly, at UTK, efforts are underway to establish a 5G testbed with contributions from AT&T. This will allow future testing of 5G C-V2X technologies.

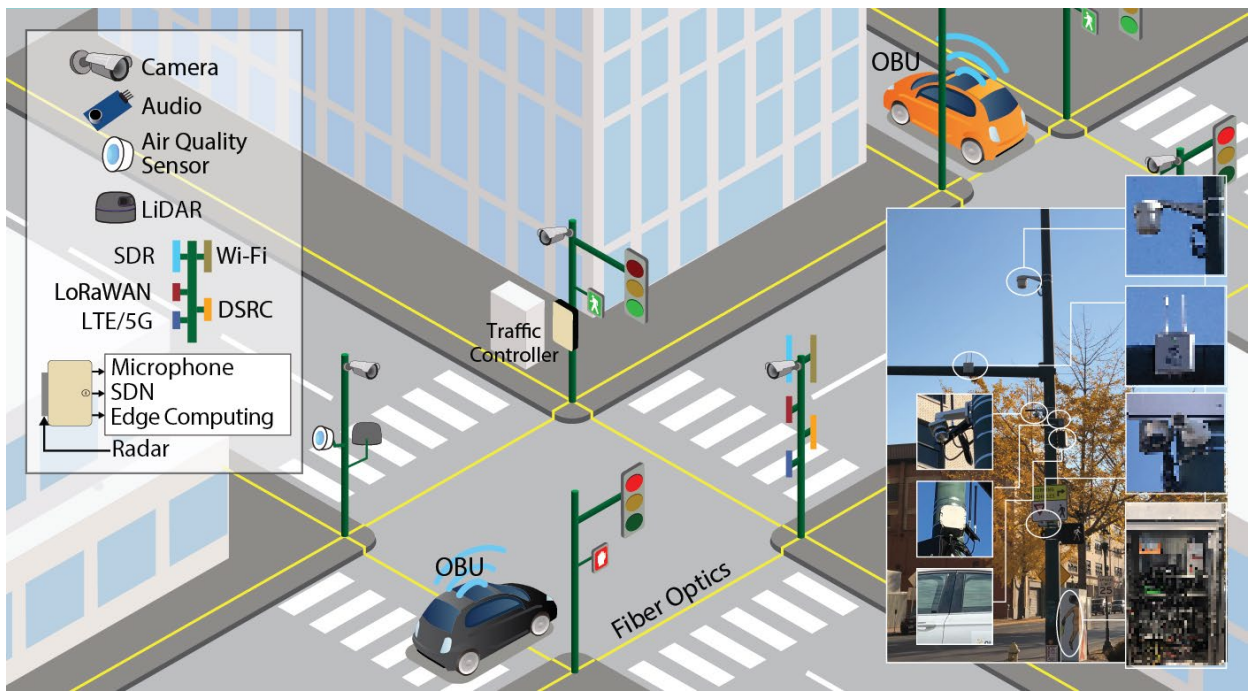

**Figure 4-1 MLK Smart Corridor Testbed in Chattanooga**

The MLK Smart Corridor is outfitted with two models of RSUs. When the testbed was first developed in 2018, Savari roadside units were installed along the corridor. An OBU was also purchased from Savari for use on the testbed. Both the OBU and RSU supported DSRC V2X communications only. Once the FCC adopted rules to repurpose the lower 45 megahertz of the 5.9 GHz spectrum, dual-mode DSRC V2X and C-V2X RSUs were procured for the testbed. Hybrid dual-mode DSRC V2X/C-V2X roadside units by vendor Applied Information, model AI-500-095, were procured in June 2021. As of February 2022, the deployment is still ongoing. Half of the units are fully operational, with the other half waiting to be configured by vendor Applied Information. A dedicated connected vehicle was purchased for the testbed in 2021, which was outfitted with an Applied Information OBU.

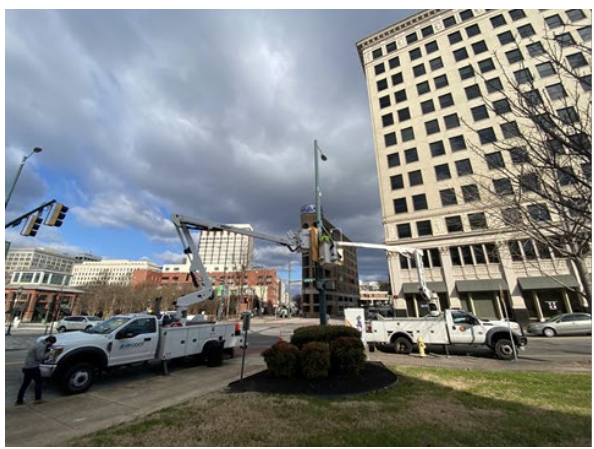 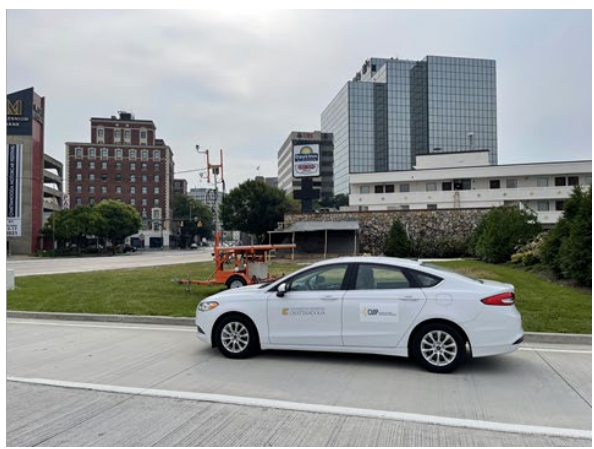

**Figure 4-2 Installation of RSU and UTC's connected vehicle**

UTC collected data along with each approach at the Martin Luther King Blvd and Georgia Avenue intersection of the corridor. SPaT messages were the data collected via the OBU by the vehicle. UTC planned to collect metrics from the network interface on the OBU to evaluate the performance. However, the OBU does not provide direct access to the network interface or the raw SPaT messages, so the default process was used for extracting messages from the OBU. Messages were written to a USB stick plugged into the OBU. The data collected represents a log of messages successfully received by the OBU. The contents of the log consist of records for each message received, including meta-data such as timestamp, radio type, message type, latitude, longitude, heading, and speed. A screenshot of the data collected on February 3, 2022, using DSRC V2X and C-V2X is shown below. This dataset is the data received by the vehicle from the RSU that includes SPaT, MAP (i.e., SAE J2735 Map Message-a MapData Message), and Traveler Information Message. UTC compared the performance of DSRC V2X and C-V2X using these data points Table 4-1.

Both DSRC V2X and C-V2X data were collected simultaneously for each round of data collection. The recorded SPaT messages on the OBU were aggregated down to the second to determine the loss/ratio or packet loss. Zero packet loss is obtained if the SPaT messages were recorded at 10 Hz, which is the rate at which they are broadcasted from the RSU. Based on UTC findings, DSRC V2X experienced a higher percentage of packet loss with a lower transmission range than C-V2X, as shown in Figure 4-3. The received packets from the RSU are plotted by latitude and longitude.

The blue dots show the locations of the vehicle where packets were received. The small red dot shows the location of the RSU at the intersection of MLK Blvd. and Georgia Ave.

The southbound approach of Georgia Avenue extends from a small side street between two buildings which reduces the line of sight with the intersection. This results in a significant reduction in range for DSRC V2X while having a less significant effect on C-V2X. The maximum range recorded in UTC data is 427 meters and 514 for DSRC V2X and C-V2X, respectively. Figure 4-3 shows a map view of the packets received by the OBU. The red marker indicates the location of the RSU, and the blue markers represent the vehicle's location when packets were received. Overall, LTE 4G seems to provide stable performance (than DSRC V2X) when considering packet loss and range in the Chattanooga urban environment.

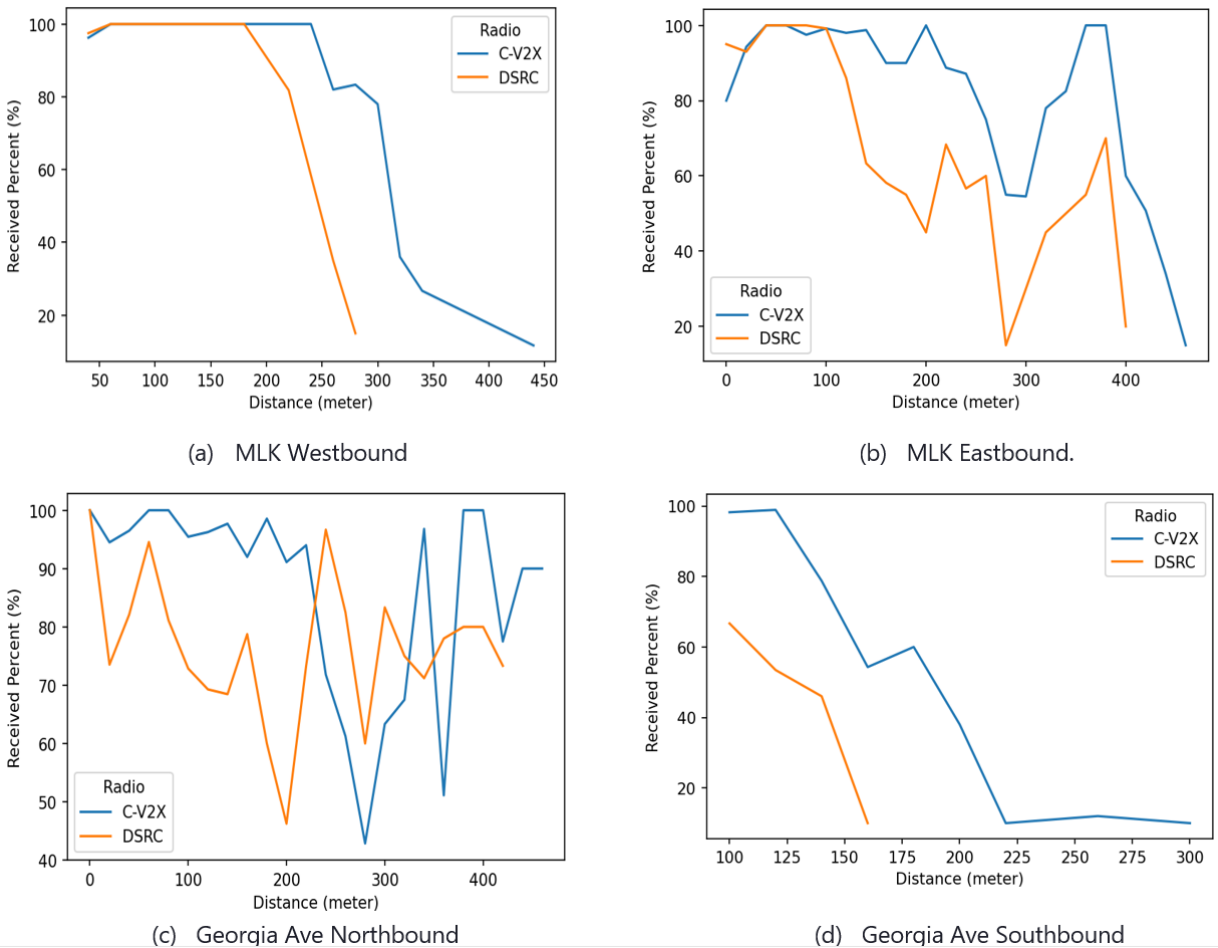

**Figure 4-3 The received percent of packets vs. distance of the vehicle from the RSU while the vehicle was approaching the intersection of the MLK and Georgia Ave in different approaches.**

| No. | Epoch Time | Local Time | Latitude | Longitude | Altitude (ft) | Heading | Speed (mph) | Radio | TX/RX | Msg Type | Transmitted | Transmitted Name |
|---|---|---|---|---|---|---|---|---|---|---|---|---|
| 1 | 1643912716 | Thu 2022-02-03 18:25:16 | 35.045135 | -85.308586 | 649.44 | 135 | 0 | DSRC | RX | MAP | 51538 | - |
| 2 | 1643912726 | Thu 2022-02-03 18:25:26 | 35.045134 | -85.308572 | 665.84 | 135 | 0.62137 | DSRC | RX | SPAT | 51538 | Georgia Ave/E M L King Blvd |
| 3 | 1643912726 | Thu 2022-02-03 18:25:26 | 35.045134 | -85.308572 | 665.84 | 135 | 0.62137 | C-V2X | RX | SPAT | 51538 | Georgia Ave/E M L King Blvd |
| 4 | 1643912726 | Thu 2022-02-03 18:25:26 | 35.045134 | -85.308572 | 665.84 | 135 | 0.62137 | DSRC | RX | SPAT | 51538 | Georgia Ave/E M L King Blvd |
| 5 | 1643912726 | Thu 2022-02-03 18:25:26 | 35.045134 | -85.308572 | 665.84 | 135 | 0.62137 | C-V2X | RX | SPAT | 51538 | Georgia Ave/E M L King Blvd |
| 6 | 1643912727 | Thu 2022-02-03 18:25:27 | 35.045134 | -85.308572 | 665.84 | 135 | 0.62137 | DSRC | RX | SPAT | 51538 | Georgia Ave/E M L King Blvd |
| 7 | 1643912727 | Thu 2022-02-03 18:25:27 | 35.045134 | -85.308572 | 665.84 | 135 | 0.62137 | C-V2X | RX | MAP | 51538 | |
| 8 | 1643912727 | Thu 2022-02-03 18:25:27 | 35.045134 | -85.308572 | 665.84 | 135 | 0.62137 | C-V2X | TIM | | - | speed limit |
| 9 | 1643912727 | Thu 2022-02-03 18:25:27 | 35.045134 | -85.308572 | 665.84 | 135 | 0.62137 | C-V2X | RX | SPAT | 51538 | Georgia Ave/E M L King Blvd |
| 10 | 1643912727 | Thu 2022-02-03 18:25:27 | 35.045134 | -85.308572 | 665.84 | 135 | 0.621371 | C-V2X | RX | SPAT | 51538 | Georgia Ave/E M L King Blvd |

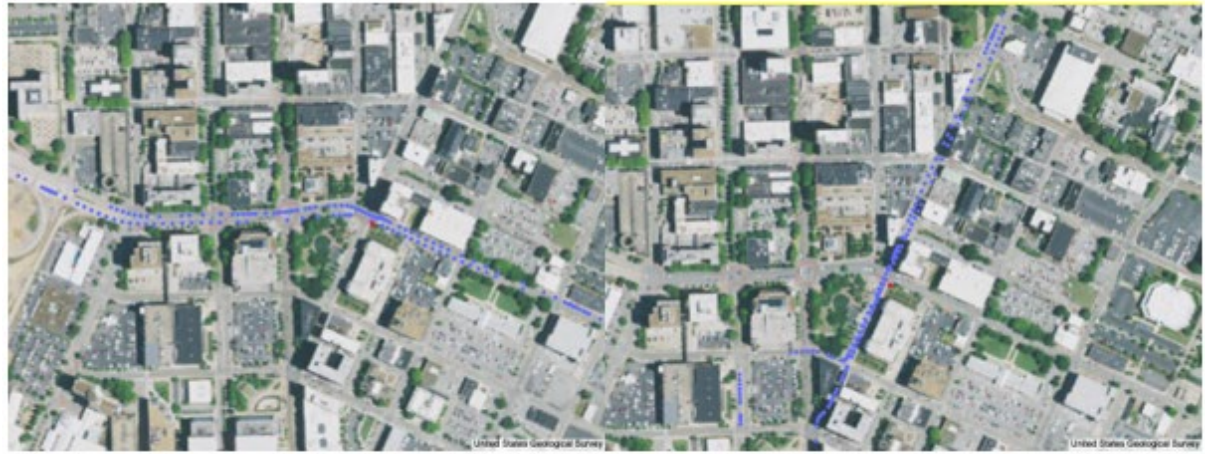

(a) Martin Luther King Blvd.          (d) Georgia Avenue

**Figure 4-4 The received packets from the RSU are plotted by latitude and longitude. The blue dots show the locations of the vehicle where packets were received. The small red dot shows the location of the RSU at the intersection of MLK Blvd. and Georgia Ave.**

# Chapter 5 CV Pilot Successes and Lessons Learned

This chapter provides details about the successes and lessons learned about communication technologies from CV pilot programs.

## 5.1 SCMS-Secure Credential Management Systems in CV Pilots

The CV pilot programs have pioneered deploying fully connected devices to Secure Credential Management Systems (SCMS), which validates CV devices to ensure trusted communication. However, the OBUs need to reach SCMS to receive additional certification, called certificate top-off. Because of the limited number of RSUs implemented in the environment for supporting the top-off service, such actions have sometimes experienced failure. Some OBUs would not be able to receive all top-off validation before leaving the RSUs coverage area. To solve this problem, the SCMS provider implemented a new device within the SCMS to record all interactions between the SCMS and other devices, which helps OBUs to download top-off certification at the appropriate times without interaction with other devices. Additionally, sometimes, SCMS experienced deactivation and could not provide certificate download. Therefore, the SCMS provider improved the deactivation configuration to enable OBUs to remain dormant for a more extended period. There were still other issues, such as failing to receive the initial responses from the SCMS by RSUs, enrollment certification expiration, and a short validity period of Pseudonym Certificate Authority. These issues resulted in some improvements, which the SCMS provider implemented. The US DOT has commenced a national SCMS development program because of the issues discussed above. The policies, procedures, and lessons learned will be leveraged as a key input to this project. This project aims to help US DOT and the industry establish a viable national SCMS ecosystem to support V2X communication. As a result, TDOT can wait until the results of this project become available, assisting in deployment while providing a more secure network for the new technology [14] [15].

## 5.2 Wireless Backhaul Issues-Tampa CV Pilot

The term backhaul is frequently used in telecommunications, and it refers to transmitting a signal from a remote site or network to another site. A connected vehicle pilot project highlighted wireless backhaul issues in Tampa, Florida. To provide a V2V and V2I communication bed, the project deployed OBUs in up to one thousand vehicles and RSUs at 47 locations along Tampa Hillsborough Expressway Authority (THEA)'s Reversible Express Lane (REL) and in the Central Business District. Additionally, as part of this project, THEA connected the RSUs to THEA's TMC via fiber. However, during the project, THEA encountered two unpredicted challenges in connecting the signal controllers and the RSUs. These challenges were delays in installing fiber in the city of Tampa and maladjustment between old pair phone lines and upgraded signal controllers using Ethernet. As a result, THEA installed cellular data modems in the RSUs to connect the controllers and the TMC. However, fiber would be the preferred means of communication and modems can serve as backups. Hence, in the future, TDOT, in partnership with IOOs, can consider installing cellular data modems in the RSUs as a backup for fiber technology, which is preferred.

## 5.3 Roadside Unit Transient Surge Immunity-Tampa CV Pilot

A vital aspect of the THEA pilot program is to resolve the technical deployment issues. During one year of continuous testing, the THEA pilot team realized that four RSUs among forty-four installed devices were not able to connect with the primary server. They started finding the reason for this problem, and finally, they found out that lightning strikes were causing damage to the RSUs. Hence, they replaced the RSUs and properly grounded them to solve this problem. It is an installation consideration that TDOT can consider.

## 5.4 New York CV Pilot Completes DSRC Licensing

The New York City Department of Transportation-led CV Pilot program has completed an 18-month FCC licensing process to use DSRC V2X to support vehicle-to-infrastructure (V2I) communication. The New York City CV pilot project area encompasses three distinct areas in the boroughs of Manhattan, Queens, and Brooklyn. The project has approximately 320 sites where CV equipment using DSRC V2X requires FCC licenses. In 2003, the FCC first established licensing and service rules for DSRC V2X for ITS in the 5.9 GHz band. According to the FCC, the ITS services on that band share co-primary status with radio services for high-powered military communications and fixed satellite communication, as well as secondary services designated for amateurs and industrial, scientific, and medical equipment. The use of a federal government-designated radio spectrum requires CV deployers to achieve a license from the FCC to operate within the band to guarantee there is no interference so that it can prevent safety concerns for the CV deployment and services such as airports.

## 5.5 New York CV Pilot Demonstration

In June 2018, multiple DSRC Aftermarket Safety Device V2V applications for the NYC connected vehicle pilot deployment were demonstrated at the 2018 ITS annual meeting. These applications included Blind Spot Warning (BSW), Electronic Emergency Brake Light (EEBL), Forward Collision Warning (FCW), and Intersection Movement Assist (IMA). Unfortunately, the selected roadway for the demonstration had thick coverage trees, so the lack of open sky interfered with Global Navigation Satellite System (GNSS) signals and made the applications unreliable. As a result, it is important that in the future, GNSS signals are improved by the deployment of RSUs to support the C-V2X technology in relatively dense coverage areas. As a result, transition to C-V2X can be coupled with GNSS signal improvements.

## 5.6 Detecting and Resolving Wireless Interference-Tampa CV Pilot

While testing the THEA CV pilot devices, Internet Protocol (IPv6) radio traffic was observed on some DSRC channels used for the CV pilot. In analyzing the captured IPv6, the identifier "HamWAN" and amateur radio were frequently observed (Figure **5-1**). Radio used in HamWAN has a range of over 50 Km, and it is also frequently used in the 5.9 GHz safety band by DSRC V2X devices. Additionally, amateur radio has a secondary allocation in the 5.9 GHz spectrum. The sample from the wireless capture device showed that the secondary use was causing harmful interference with DSRC V2X. As a result, the THEA team has decided to continuously monitor for potential interference through random wireless monitoring (sniffer). The team also recommended that other CV deployers implement an early and complete radio spectral analysis for deployment areas to prevent interference.

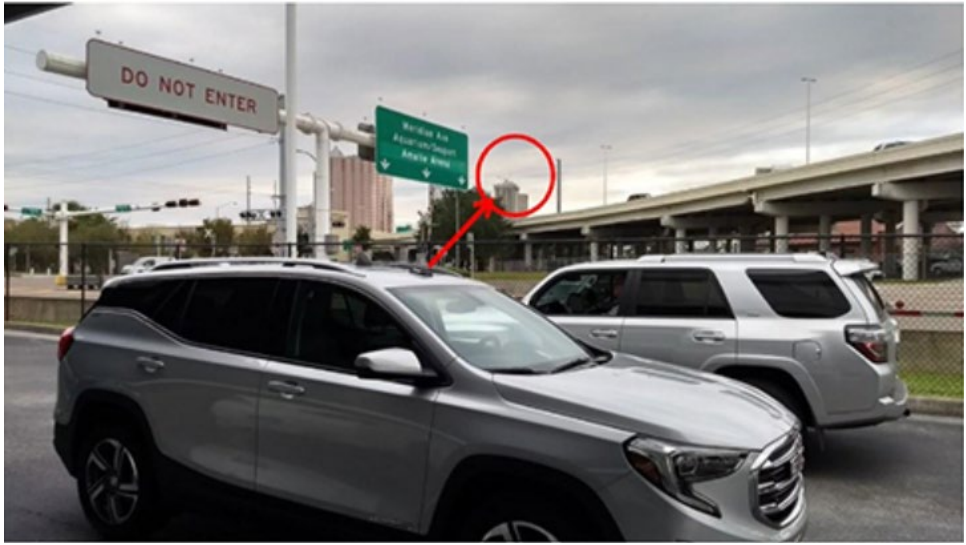

**Figure 5-1 The position of the vehicle and OBU antenna on that as well as the line of sight to the HamWAN radio antenna**

**Source:** https://www.its.dot.gov/pilots/thea_wireless_interference.htm

## 5.7 Ambiguity in Communication Standard in Three CV Pilot Sites

There are several channels within the 5.9 GHz safety band, and Channel 172 is the primary channel that carries safety-related information. The New York and Tampa CV pilots have decided to use dual radios in vehicles. However, one listens to channel 172, and the other listens to another channel for complementary information, e.g., Traveler Information Message. On the other hand, the Wyoming CV pilot has decided initially to use a single radio and listen to channel 172 and switch to other channels in time of need. However, switching away from the channel is not safe due to some important safety messages that should be lost. As a result, the Wyoming team has accepted re-designing its communication beds to include dual radios.

## 5.8 Recruiting Volunteers for CV tests-Tampa CV Pilot

The Tampa CV pilot has recruited volunteers from the general public. Specifically, THEA offers monetary incentives to volunteers from the general public, e.g., the Tampa CV Pilot offered toll savings of up to $550 to drivers of Honda, Acura, Toyota, Kia, or Hyundai vehicles. The vehicles were installed with OBUs that included antennas, a DSCRC radio device, and a heads-up display at service stations. A survey assessed the eligibility of drivers and their vehicles. Moreover, directions were given to recruits about the installation of CV devices and what to expect when driving. A sampling of the information provided is available here:

https://theacvpilot.com/participants/acura/

## 5.9 CAV Data Collection, Management, and Use

Given the large amounts of CAV data, it is critical to have a data management plan (DMP), which describes the CAV and related data to be collected and managed through the duration of a test. The DMP also defines a framework for sharing the data with US DOT and independent evaluators. Such plans ensure that the data are available for use in performance measurement, and through the process, data privacy is assured. More information is available in supplementary project report number 4, "Ecosystem for Connected and Automated Vehicles: Investments in Data Collection, Analytics, and Simulations" [16].

More resources about data collection, sharing, and management are at:

https://www.its.dot.gov/pilots/technical_assistance_events.htm

# Chapter 6  Illustrative Practices in Other States

Case studies in this section are illustrative of how states are procuring RSUs and OBUs from vendors.

## Case study 1: Colorado

Colorado plans to procure a substantial number of C-V2X devices but first will test a few devices. The Request for Proposals (RFP) for their purchase is illustrative [17]**.**

> - *Colorado plans to procure a substantial number of C-V2X devices, but first, it will test a few devices.*
> - *Agencies are examining strategies to mitigate the risk by using dual radio DSRC V2X and C-V2X radio.*

**Colorado DOT's recent RFP for RSUs and OBUs**

The Colorado Department of Transportation (CDOT)'s innovation office implements the State's CV strategy and deployment. CV systems use the 5.9 GHz frequency band for wireless communication, enabling data exchange between vehicles and infrastructure. CDOT has provided a plan for future infrastructure expansion, including the potential real-time data exchangeability enabled by CV technology. Note that in some locations, the town manages some of the traffic signals CDOT also manages other traffic signals through partnerships between the local and state agencies. Thus, CDOT owns and operates some of the traffic signals in Colorado. The programs include the improvement of the CV ecosystem that stores and analyzes CDOT's CV data. In order to store the data, the state and local agencies typically have a plan for building new data centers similar to THEA CV Pilot in Tampa, Florida. CDOT expects that the project will be funded in part through federal funds. Accordingly, applicable federal law and regulations will apply to the project's procurement and the resulting project agreement. This issue is addressed further in project report number 4 [18].

**Project Goals**

CDOT's objective for the RFP is to detect qualified vendors to supply RSUs and OBUs to CDOT on a defined timeline. CDOT plans to implement 1000 RSUs and 500 OBUs purchased from qualified manufacturers over the contract's lifetime. CDOT will test OBUs and RSUs to examine their quality, including post-award acceptance, validation, and Firmware revision test. Acceptance testing will be implemented in a bench environment before the deployment, and validation tests will also be conducted in bench and field environments. A firmware revision test will be done for all firmware updates.

Additionally, the selected vendors will conduct the physical installation of OBUs and RSUs. The RSUs and OBUs selected should have the ability to enroll in CDOT's SCMS as CDOT owns a CV SCMS. Contractors must provide the goods/services shown in TABLE **6-1**.

| Goods/Services | Estimated Delivery Date |
|---|---|
| Technical consulting and project management<br><br>Project management plan | Continuous over the whole device life cycle |
| Delivery of initial 75 CV RSUs and 30 CV OBUs | Within 60 days |
| Delivery of an additional 150 CV RSUs and 45 CV OBUs | Within 120 days |

**Documentation requested by CDOT from vendors**

1. The selected vendors should provide any supporting material, including user manuals and reference guides, to assist CDOT in configuring, deploying, and maintaining CV RSUs and OBUs.
2. The manufacturer should provide documentation for the Omni air certification testing for CV2X and RSU specification tests.
3. Finally, the selected proposer should provide firmware upgrades and revisions at any time when the former documentations become obsolete.

## Case study 2: Georgia

To hasten the deployment of V2X technology, the Georgia Department of Transportation (GDOT) has used public-private-philanthropic partnerships (not presently allowed in Tennessee) to build smart infrastructure. GDOT has developed V2X infrastructure along I-85 in West Georgia. Initially, the project focused on an 18-mile corridor of the rural interstate by deploying a few C-V2X devices. Subsequently, the project's scope was expanded by adding seven new radios and ten additional C-V2X devices. Georgia-based partners are examining strategies to mitigate the risk associated with communication devices by using dual-mode DSRC V2X and C-V2X radio.

For example, the vendor "Applied Information" offers the dual-mode CV system, which enables communication with both 4G LTE C-V2X and DSRC V2X. Additionally, the equipment can be upgraded to 5G (through an open modem slot). "Applied Information" provides relevant services in this case, while other vendors such as Commsignia also provide similar products. One of the items in this early stage of the technology is to modularize the hardware design of the radio. TDOT can consider requesting information about the two radios embedded in the RSU module. TDOT can consider such RSUs after ensuring that performance is adequate and the quality of the equipment is appropriate.

# Chapter 7  Conclusions and Recommendations

This study supports TDOT decisions on communications technologies. The research team has compared standards for vehicle communications (DSRC V2X and C-V2X technologies) and explored the transition from DSRC V2X technology to C-V2X given the recent safety band ruling by the FCC. As C-V2X communication technology has not been tested at a large scale, the technology currently has substantial uncertainty, which can be hedged by using dual-mode C-V2X/DSRC V2X devices. The key findings from a review of recent US DOT guidance on vehicular communication and the literature highlight the following issues.

*Based on the findings, recommendations can be categorized as:*
- *Managing transition to dual-mode C-V2X and DSRC V2X devices*
- *Managing and harnessing CAV data*
- *Conducting research on communications technologies*

- ***Transition to C-V2X technologies.*** Agencies throughout the US that have deployed DSRC V2X are now transitioning to C-V2X technologies, which nevertheless involves uncertainty given the complexities of procuring and installing the software and hardware. Furthermore, technology, equipment, standards, and uses for C-V2X are evolving and agencies have found the transition from DSRC V2X to C-V2X challenging. While the transition undertaken by TDOT can be done by working with vendors, this involves more than swapping DSRC V2X devices with LTE C-V2X devices or installing dual-mode devices that can perform both DSRC V2X and C-V2X communications. Notably, these devices and associated software are still in their infancy, have limited availability through vendors, and entail several complicated steps, with associated uncertainties, according to a recent National Cooperative Highway Research Program report [1]. This transition will involve more time and resources.

- ***Issues in C-V2X transition.*** Complicating the migration to C-V2X is TDOT's role in traffic signal operations and maintenance, which is limited to funding and designing/constructing traffic signals. However, local agencies' infrastructure owners and operators currently operate and maintain signals, and TDOT works and assists local agencies in the smooth operation of signals. Hence local agencies will work with TDOT to deploy, operate, and maintain C-V2X technology. Further complications come from LTE C-V2X interference from unlicensed devices and channel congestion, adversely affecting safety-critical applications. Since there are substantial uncertainties in transitioning to these emerging technologies, discussions with IOOs about the operation and maintenance of C-V2X may have to wait to resolve these issues. At the same time, TDOT can invest in limited experimentation with dual-mode devices. Since there are substantial uncertainties in transitioning to these emerging technologies, IOOs in Tennessee can examine the operation and maintenance of dual-mode C-V2X. Notably, TDOT has recently issued guidance recommending a transition to dual-mode C-V2X devices if IOOs are in planning or design stages. Meanwhile, it will be prudent to develop investment plans to experiment with dual-mode C-V2X devices.

- **Smart infrastructure deployment.** In Tennessee, smart infrastructure deployment will require a successful transition from DSRC V2X to C-V2X communications, especially with RSUs, e.g., installed at traffic signals or on freeways. This transition has a high opportunity cost for changing communication devices and delaying the deployment of safety-critical applications through the available DSRC V2X devices. Specifically, researchers at the University of Michigan Transportation Research Institute [2] point out that the cost of delaying the deployment of safety-critical applications through the available DSRC V2X can be measured in terms of tens of thousands of lives lost over five years.

- **Connected vehicle applications and needed research.** Connected vehicle applications in Tennessee have relied on the DSRC V2X communication platform for Transit Signal Priority and Emergency Vehicle Preemption applications. CAV data collection, use, security, and storage are critical aspects going forward. Notably, 4G LTE-based C-V2X communication technology is being tested by the USDOT. Early results of data analysis indicate that there can be congestion issues during operation compared with DSRC V2X. However, presently there is not enough evidence to confirm or contradict whether LTE C-V2X will scale up to the safety issues. More research is needed on several aspects of communication technology. While the FCC has provided a transition plan for moving from DSRC V2X into the new C-V2X spectrum, the shift will nevertheless be costly.

The recommendations regarding V2X communications are structured around three main themes, the transition, data, and research/testing:

- **Managing the transition to dual-mode C-V2X and DSRC V2X devices.** Given that C-V2X technology is still being examined for large-scale deployment, the transition will include testing and using dual-mode DSRC V2X and C-V2X radio devices. Furthermore, strategies can include investing in the reliable installation of software updates for C-V2X (e.g., updates that address security vulnerabilities within signal controllers, so they can communicate securely with C-V2X devices in the future); recruiting volunteers from the general public and/or fleet vehicles (owned by TDOT, or other agencies) for installing onboard units; and exploring automotive/OEM industry's acceptance of C-V2X technologies. Importantly, experimentation with dual-mode devices by UTC has shown promising results for dual-mode devices. Specifically, experimentation in the Chattanooga MLK smart corridor shows that LTE 4G provides stable performance (than DSRC V2X) when considering packet loss and range in an urban environment.

- **Managing and harnessing connected and automated vehicle data.** It is important to identify CAV data needs and types (e.g., basic safety messages, SPaT messages, and vehicle trajectories). Equally important are data analytics and cybersecurity investments for streaming data generated by CAVs. Data management plans are also needed. Harnessing the data is an opportunity to improve traffic operations and performance evaluation. TDOT, with private sector partners, can collect high-resolution data from traffic signals and vehicle probe data, allowing engineers to track traffic trends at an intersection, corridor, and programmatic level as well as visualize maintenance problems. In this regard, TDOT can use Automated Traffic Signal Performance Measures (ATSPM) software to oversee real-time and historic functionality at individual intersections. All intersections communicating with TDOT can be configured in ATSPM. ATSPM can find faults and errors, thereby saving staff field inspection time. This software has several use cases, including Utah DOT, Virginia DOT, and Georgia DOT. In addition, TDOT can consider using software named TEAMS (Traffic Engineering and Asset Management Software) to track traffic signal equipment, including signs, Americans with Disabilities (ADA) compliance, and overall rating of intersections. For equipment maintenance updates, if the equipment is not performing correctly, it is noted as a task in TEAMS and updated as needed.

- ***Future research on communication technologies.*** Future research can entail testing the performance in terms of latency and packet drop of C-V2X technologies and the impacts of (user) congestion and interference in communication networks. Also, assessing TDOT's tolerance for latency and packet loss in safety-critical applications is important. Furthermore, investments in addressing the uncertainty of C-V2X in terms of the scale (number of vehicles) will be valuable, along with testing the longevity and compatibility of C-V2X technology performance in different environmental conditions within Tennessee. Noting that the current 30 MHz for communication may not be sufficient, there is a need to investigate how the 5.9 GHz communication band can support C-V2X. More generally, for TDOT to deploy C-V2X technologies, transportation system readiness should be assessed. Overall, substantial and comprehensive testing and research are needed before fully transitioning to C-V2X.

# Appendix A: US DOT LTE-V2X Radio Performance Tests

The US DOT test methodology includes laboratory device performance characterization, device/system integration testing, controlled track testing, and small-scale real-world testing. U.S. DOT implemented three scenarios (described below) to examine the LTE-V2X radio performance in safety-critical transportation conditions. In these safety-critical scenarios, the presence of DSRC V2X devices was considered to identify the interference between these two technologies operating simultaneously. The study is also finding solutions to prevent interference in the future. DSRC V2X and C-V2X test parameters for all three scenarios are shown in Table A-1.

TABLE A- 1 TEST PARAMETERS (5)

| Devices | Test parameters |
|---|---|
| Number of OBUs | 250+ |
| Number of RSUs | 10+ |
| Payload | 365 Byte (V2V/1400 Byte (I2V) |
| DSRC V2X Transmit Power | 20 dBm |
| DSRC V2X High-speed Vehicle Congestion Mitigation | ON |
| DSRC V2X High-speed Vehicle TTI Setting | AUTOMATIC |
| DSRC V2X Channels | 172/180/184 |
| LTE-V2X Channel for 250+ | 183 |
| LTE-V2X 250+ Stationary Devices Congestion Mitigation | ON |
| 250+ Stationary OBU Devices TTI Setting | AUTOMATIC |
| Obstructing Vehicle | YES/NO |

**Scenario 1:** High-speed moving vehicles with Non-line-of Sight in the midst of congestion, communicating with each other and with surrounding static vehicles

This scenario aims to evaluate the communication among vehicles operating at high speed (over 75 mph) with Non-line-of Sight (NLOS). At the same time, OBUs are stationary as the LTE-V2X technology uses different modulation in relation to vehicles' speed, which is demonstrated in Figure A-1. In Scenario 1, the LTE-V2X test parameters are shown in Table A-2.

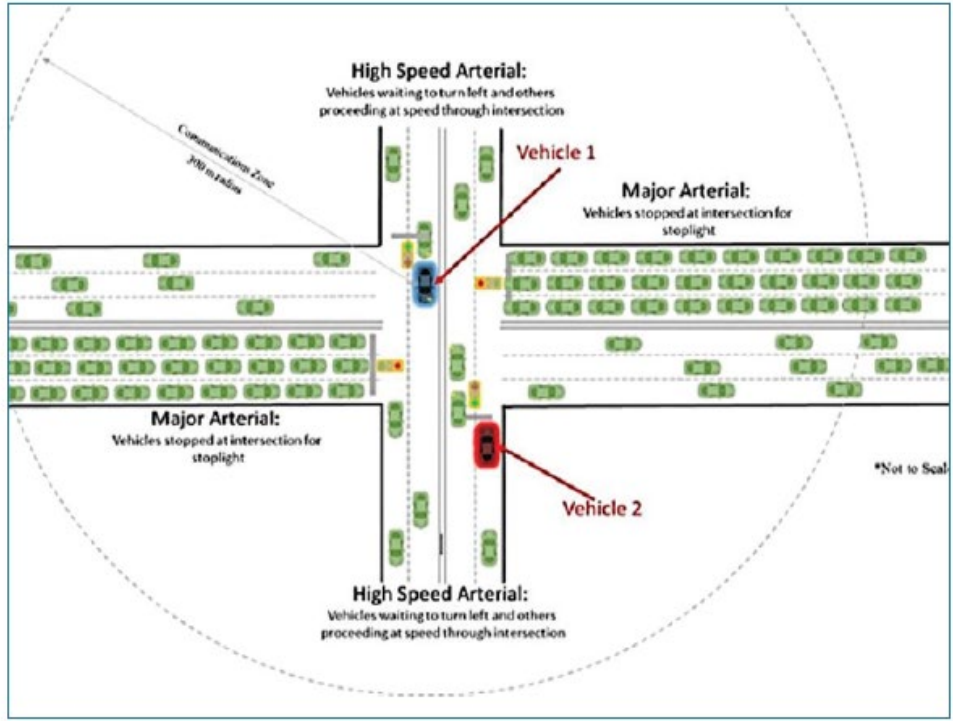

**Figure A- 1 Test Track Scenario 1 Set-up [9]**

TABLE A- 2 SCENARIO 1 LTE-V2X TEST PARAMETERS [9]

| *LTE-V2X test parameters* | *parameters* |
|---|---|
| *Number of OBUs* | 250+ |
| *Number of RSUs* | 10+ |
| *Payload* | 365 Byte (V2V/1400 Byte (I2V) |
| *Hybrid Automatic Repeat Request (HARQ)* | ON |
| *LTE-V2X Channel* | 183 (for both OBUs and RSUs) |
| *LTE-V2X Transmit Power* | 20 dBm |
| *Obstructing Vehicle* | YES/NO |
| *High-speed Vehicle Congestion Mitigation* | ON/OFF |
| *High-speed Vehicle Transmit Time Interval (TTI) Setting* | AUTOMATIC/100 ms |
| *250+ Stationary Devices Congestion Mitigation* | YES/NO |
| *250+ Stationary Devices TTI Setting* | AUTOMATIC/100 ms/300 ms/600 ms |

**Scenario 2:** High-Speed Moving vehicles during Congestion, Communicating with each other and with surrounding Static vehicles

This scenario examines communication between high-speed vehicles and surrounding stationary vehicles in congested areas with and without unlicensed Wi-Fi transmissions, which is illustrated in Figure A-2. This scenario is implemented to evaluate the suitability of LTE-CV2X communications to support crash-imminent safety applications. Scenario 2 LTE-CV2X test parameters are shown in Table A-3.

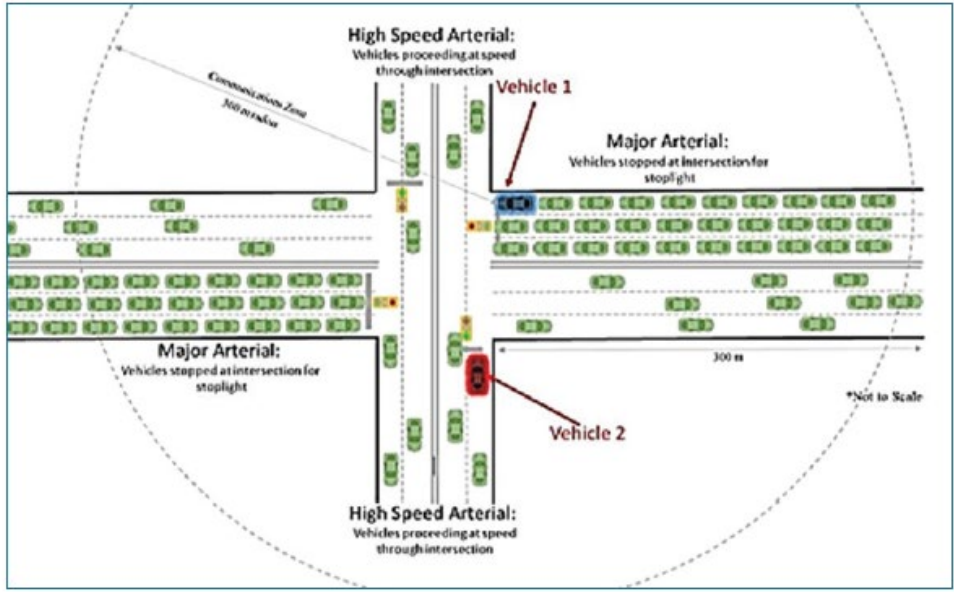

**Figure A- 2 Test Track Scenario 2 Set-up [9]**

TABLE A- 3 SCENARIO 2 LTE-V2X TEST PARAMETERS [9]

| LTE-V2X test | parameters |
|---|---|
| Number of LTE-V2X OBUs | 250+ |
| Number of LTE-V2X RSUs | 10+ |
| Payload | 365 Byte (V2V)/1400 Byte (I2V) |
| Hybrid Automatic Repeat Request (HARQ) | ON |
| LTE-V2X Channel | 5.915 MHz 183 (for both OBUs and RSUs) |
| LTE-V2X Transmit Power | 20 dBm |
| High-speed Vehicle Congestion Mitigation | ON/OFF |
| High-speed Vehicle Transmit Time Interval (TTI) | AUTOMATIC |
| 250+ Stationary Devices Congestion Mitigation | YES/NO |
| 250+ Stationary Devices TTI Setting | AUTOMATIC/50 ms/100 ms/600 ms |
| Unlicensed WI-FI Effective Isotropic Radiated Power (EIRP) | OFF/13dBm/33dBm/Max OOBE |

**Scenario 3:** Testing in a real-world environment

The goal of the scenario is to use a CV pilot site and a real-world location for communications performance testing in different conditions (different weather and times of day), as shown in Figure A-3. Scenario 3 LTE-V2X test parameters are shown in Table A-4.

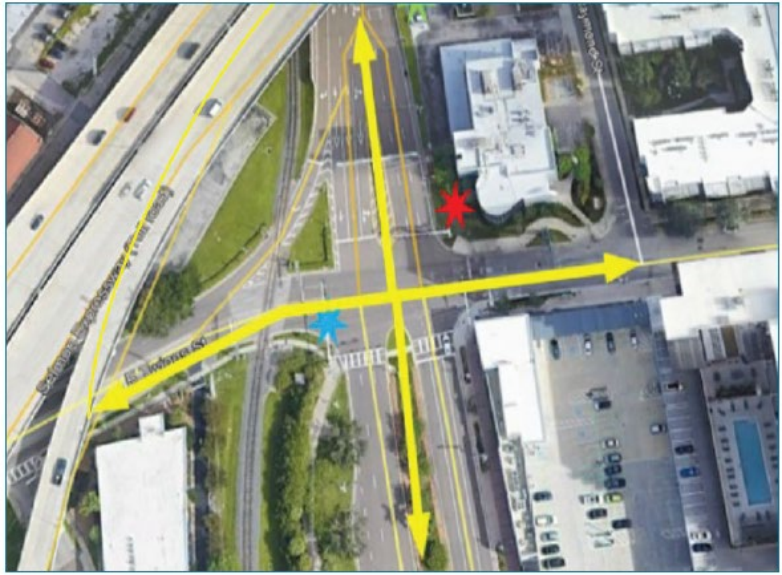

**Figure A- 3 Test Track Scenario 3 Set-up [9]**

| LTE-V2X test | Parameters |
|---|---|
| Number of LTE-V2X OBUs / RSUs | 3 OBUs, 1 RSU |
| Payload | 365 Byte (V2V)/1400 Byte (I2V) |
| Hybrid Automatic Repeat Request (HARQ) | ON |
| LTE-V2X Channel | CH 183 |
| LTE-V2X Transmit Power | 20 dBm |
| Unlicensed WI-FI Channel | 20 MHz channel centered at 5.895 MHz/ Also in the next adjacent channel CH177 Unlicensed WI-FI |
| Unlicensed WI-FI EIRP | OFF/13dBm/33dBm/Max OOBE |

By analyzing data from scenarios, the following issues can be addressed:

- LTE-V2X device performance: Whether LTE-V2X devices remain operational in conditions when GPS is lost? Do they lose their ability to synchronize with surrounding devices?
- Congestion: Whether LTE-V2X devices maintain their performance in congested areas of the digital network?
- Interference: Are mitigations required to prevent harmful interference between DSRC V2X and LTE-V2X devices? This includes interference from unlicensed Wi-Fi OOBE above and below the Safety Band.
- Safety: What is the potential for LTE-2X radio performance to overcome device crashes?

# Appendix B: Experience in Utilizing Dual-mode DSRC V2X and C-V2X Devices

***THEA Connected Vehicle Pilot***

"As one of three initial locations selected nationally by the United States Department of Transportation (USDOT), the Tampa Hillsborough Expressway Authority (THEA) has been working since 2018 to deploy and pilot Connected Vehicle (CV) applications to demonstrate safety and mobility benefits of the technology in and around downtown Tampa.

In late 2020, the THEA CV Pilot took an exciting next step in progressing the technology from a pilot toward more mainstream deployment. The current phase of the THEA CV Pilot builds on the project's existing deployment of Dedicated Short-Range Communications technology by deploying Roadside Units (RSUs) with dual-mode DSRC V2X and Cellular Vehicle to Everything (C-V2X) capabilities. In addition, the project includes the field deployment of testing equipment to generate DSRC V2X, C-V2X, Wi-Fi, and other radio signals. The functionality of these devices, combined with the dual-mode RSUs and the pilot's existing and expanding CV Onboard Unit (OBUs) deployment, are being implemented to enable radio spectrum interference and channel capacity testing in a live, real-world operating environment. The testing will be performed in collaboration with ongoing USDOT Intelligent Transportation Systems (ITS) Joint Program Office (JPO) research and is facilitating a more thorough understanding of the impacts of a joint CV technology environment that is anticipated with the expected finalization of the Federal Communications Commission's (FCCs) changes to the spectrum allocation in the transportation "safety band"."

For additional information, please visit the website:

https://www.fdot.gov/traffic/its/projects-deploy/cv/maplocations/thea-cvp.shtm

www.its.dot.gov/pilots/pilots_thea.htm

***Georgia Department of Transportation CV Project***

Georgia is moving forward with the second phase of a connected vehicle project, expanding the area covered and partnering with carmaker Kia to outfit Georgia Department of Transportation (GDOT) vehicles with vehicle-to-everything (V2X) technology. The project, which uses Panasonic technology on an 18-mile stretch of Interstate 85, known as The Ray, will add seven new roadside communications radio units and ten additional connected vehicles.

"The innovative public-private-philanthropic partnership began in 2019 to create and install a digital testing environment focused on critical interstate use cases, such as crash and weather warnings, for stakeholder engagement and education. The first phase of work focused on an 18-mile corridor of rural interstate, known as The Ray Highway. It established a connected vehicle ecosystem with six dual-mode and dual-active roadside radios, several cellular V2X (C-V2X) GDOT connected vehicles, and the CIRRUS by Panasonic cloud-based data management platform. The dual-mode and dual-active features of the roadside radios make the infrastructure interoperable and able to communicate directly with connected vehicles through either dedicated short-range communication V2X or cellular V2X communication protocols. The new project phase expands the partnership and nearly triples the project's breadth by adding seven new radios and ten additional C-V2X connected vehicles provided by Kia Georgia. The information delivered from the dual-active radios, called "traveler information messages" (TIMs), will be delivered and displayed into the vehicles with a heads-up display (HUD) that reduces driver distraction."

For additional information, please visit the website:

https://theray.org/2021/06/28/georgias-v2x-connected-interstate-adds-kia-prepares-for-connected-freight-operations/

https://theiatl.com/projects/

### Hawaii DOT CV Pilot

The state of Hawaii is moving forward with dual-mode communication devices. "HONOLULU – The Hawaii Department of Transportation (HDOT) announces the availability of connected vehicle technology within the Ala Moana Boulevard / Nimitz Highway corridor. The connected vehicle technology can provide alerts and other information about the corridor to motorists, bicyclists, and pedestrians through a free application.

Roughly 720 detection zones (cameras and pucks) and 34 roadside units were installed through the Nimitz V2E Pilot. Communications between the detection zones, roadside units/controllers, and the Advanced Traffic Management System (ATMS) is facilitated through secure C-V2X (cellular) and DSRC V2X signals.

These roadside units are equipped with Qualcomm chipsets for both DSRC V2X and C-V2X communication, which assists in meeting the SPaT challenge criteria while ensuring that intersections are equipped with both forms of communication."

For additional information, please visit the website/document:

https://hidot.hawaii.gov/blog/2020/08/06/hdot-launches-connected-vehicle-pilot-on-ala-moana-boulevard-nimitz-highway/

https://hidot.hawaii.gov/highways/files/2020/08/V2X-Enabled-TCS-Innovations-Proposal-Econolite-Systems-redacted.pdf

### Alabama DOT CV Pilot

The state of Alabama is moving forward with dual-mode communication devices. DSRC V2X Radio and Cellular Connected Vehicle Technology is being installed in Tuscaloosa, AL, and Northport, AL. "The University of Alabama, in conjunction with the Alabama Department of Transportation, deployed connected vehicle technology at 85 major intersections in Tuscaloosa, AL, and Northport, AL. Both dedicated short-range communication V2X radio and cellular (C-V2X) connected vehicle technologies were deployed and installed to compare the effectiveness of the two technologies."

For additional information, please visit the document:

https://appinfoinc.com/wp-content/uploads/2020/06/dsrc-cellular-tuscaloosa-al.pdf

# Appendix C: Resources for Safety Band and CAV Data

Repository of USDOT info related to FCC's decision regarding Safety Band
https://www.transportation.gov/content/safety-band

Testing LTE-CV2X Radio Performance
https://www.transportation.gov/V2XTestSummary

CV Deployer Resources from USODT Turner-Fairbank Research Lab & USDOT ITS Joint Program Office
https://www.pcb.its.dot.gov/CV_deployer_resources.aspx

Colorado DOT's recent RFP for RSU and OBU
https://www.bidnetdirect.com/colorado/colorado-department-of-transportation/solicitations/Connected-Vehicle-Roadside-Unit-RSU-and-Onboard-Unit-OBU/0000270113?purchasingGroupId=8409951&origin=2

CV Pilot Success Stories and Lessons Learned
https://www.its.dot.gov/pilots/success_lessonslearned.htm

ITS Data Hub – Publicly available CV data repository
https://www.its.dot.gov/data/index.htm

CV Pilot-Illustrative Data Management Plans
https://www.its.dot.gov/pilots/technical_assistance_events.htm

FCC, First Report and Order, Further Notice of Proposed Rulemaking, and Order of Proposed Modification in the Matter of Use of the 5.850-5.925 GHz Band, Federal Communications Commission, Washington, DC, 2020.

https://docs.fcc.gov/public/attachments/FCC-20-164A1.pdf

National Academies of Sciences, Engineering, and Medicine, "Evaluation and Synthesis of Connected Vehicle Communication," The National Academic Press, Washington, DC, 2021
https://www.nap.edu/catalog/26370/evaluation-and-synthesis-of-connected-vehicle-communication-technologies

Sayer, J., C. Flannagan, & A. Leslie, The Cost in Fatalities, Injuries and Crashes Associated with Waiting to Deploy Vehicle-to-Vehicle Communication, University of Michigan Transportation Research Institute Ann Arbor, Michigan, USA,

https://deepblue.lib.umich.edu/bitstream/handle/2027.42/147434/The%20Cost%20Associated%20with%20Waiting%20to%20Deploy%20DSRC%20032018.pdf%3Fsequence=1&isAllowed=y.